\documentclass[aps,prb,groupedaddress, floatfix,showpacs, twocolumn]{revtex4-1}
\usepackage{graphicx}
\usepackage{dcolumn}
\usepackage{bm}
\usepackage{color}
\begin{document}

\title{Entanglement Measures for Quasi-Two-Dimensional Fractional Quantum Hall States}

\author{ J. Biddle$^1$, Michael R. Peterson$^{1,2}$,  and S. Das Sarma$^1$} 
\affiliation{$^1$Condensed Matter Theory Center, Department of Physics, University of Maryland, College Park, Maryland 20742, USA}
\affiliation{$^2$Department of Physics, University of California, Santa Barbara, California 93106, USA}
\date{\today}

\begin{abstract}
We theoretically examine entanglement in fractional quantum hall states, explicitly taking into account and emphasizing the quasi-two-dimensional nature of experimental quantum Hall systems.  In particular, we study the entanglement entropy and the entanglement spectrum as a function of the finite layer thickness $d$ of the quasi-two-dimensional system for a number of filling fractions $\nu$ in the lowest and the second Landau levels: $\nu$ = 1/3, 7/3, 1/2, and 5/2.  We observe that the entanglement measures are dependent on which Landau level the electrons fractionally occupy, and find that filling fractions 1/3 and 7/3, which are considered to be Laughlin states, weaken with $d$ in the lowest Landau level ($\nu$=1/3) and strengthen with $d$ in the second Landau level ($\nu$=7/3).  For the enigmatic even-denominator $\nu=5/2$ state, we find that entanglement in the ground state is consistent with that of the non-Abelian Moore-Read Pfaffian state at an optimal thickness $d$.  We also find that the single-layer $\nu = 1/2$ system is not a fractional quantum Hall state consistent with the experimental observation.  In general, our theoretical findings based on entanglement considerations are completely consistent with the results based on wavefunction overlap calculations.
\end{abstract}

\maketitle
\section{Introduction}

The discovery of the fractional quantum hall effect (FQHE) in 1982\cite{TSG1982} has proven to be one of the most significant experimental findings in all of physics. 
The incompressible quantum fluid that manifests in a two-dimensional (2D) electron system in the presence of a strong perpendicular magnetic field at low temperatures can not be explained by the ``conventional" Landau theory of phase transitions\cite{Laughlin83}. Instead this unique phase is an example of an emergent topological state of matter, where the charged quasi-particle excitations are governed by anyonic rather than fermionic or bosonic statistics\cite{Halperin84, Leinass77, Wilczek82, Arovas}.  The unique nature of FQHE becomes obvious with the realization that the phenomenon occurs only in the truly strongly interacting limit of vanishing (or extremely small) kinetic energy with the noninteracting ground state having a macroscopically large degeneracy.  Although the FQHE was discovered decades ago\cite{sarma1997perspectives, Girvin:1987a, jain2007composite}, many questions still linger to this day.  The most notable of these is the nature of the FQH state observed at electron filling fraction $\nu = 5/2$.  First discovered in 1987\cite{Willett87}, the $\nu=5/2$ state is (so far) the only exception to the famous ``odd denominator" rule for monolayer FQH systems given by the Laughlin ansatz\cite{Laughlin83} and the more general ``composite fermion" theory\cite{Jain89, jain2007composite}.  (The odd denominator rule assumes FQHE in a monolayer system.  Bilayer FQH systems, however, have several even denominator states that are well-understood.  See, for example Refs.~\onlinecite{Suen92, Eisen92, SongHe93}).  Currently, the leading theoretical candidate for the $\nu=5/2$ state is the Moore-Read (MR) Pfaffian model state introduced by Moore and Read in 1990\cite{MooreRead91}. Recently it has been noted that the particle-hole conjugate to the MR Pfaffian, the so-called anti-Pfaffian ~\cite{Lee07, Levin07}, is also a viable candidate for the FQHE at $\nu=5/2$.   An exceptional feature of these ansatz is the prediction of  anyonic quasi-particles with non-Abelian braiding statistics~\cite{MooreRead91,Nayak1996,BondersonGurarieNayak}.  This prediction is attractive given the current interest in topological quantum computing~\cite{DasSarmaTQC05,Nayak08}, but the true nature of the FQH $\nu=5/2$ state is, by no means, a settled question\cite{StorniPRL,Storni-Morf, Wojs}. Directly probing the topological nature of FQH states has proven to be both a theoretical and an experimental challenge, making definitive and uncontroversial verification of the MR Pfaffian ansatz elusive.  However, recent developments in the field of quantum information have shown that measures of entanglement are useful tools in examining the global features of many-body strongly correlated quantum states\cite{Amico08}.  

The true nature of the $\nu=5/2$ FQHE state is one of the most prominent open questions in condensed matter physics, and is a primary motivation for the current study.  Recent experimental studies have explored this mysterious state and give some weight to the Moore-Read theory that is believed to explain it.
One of these studies is the recent experiment by Venkatachalam \textit{et. al.}~\cite{LocalCharge} that measured the charge of localized excitations in the $\nu=5/2$ state to be $e/4$ as predicted by the MR theory\cite{MooreRead91}.  These results are consistent with previous studies by Radu \textit{et. al.} and Dolev \textit{et. al.} that used shot noise to investigate the local charge~\cite{Dolev08,Radu,Dolev2010}.
In addition to the experiments noted above, Willett \textit{et. al.}~\cite{WillettPNAS09, Willett09} have seen evidence of quasiparticle interference oscillations that support the existence of charge $e/4$ excitations at $\nu=5/2$.
Another recent study is work performed by Bid \textit{et. al.}\cite{NeutralMode} that experimentally observed the theorized neutral mode of the $\nu=5/2$ state consistent with the MR theory\cite{Milovanovi, Overbosch}.  Although these developments point to the MR theory as the likely candidate for the $\nu=5/2$ state, they are not sufficient to unambiguously establish the existence of non-Abelian anyons.  Also, it should be noted that the MR theory predicts a spin polarized state at $\nu=5/2$, but recent experimental work~\cite{Rhone2011, Stern2010} suggest that this state may be unpolarized in some cases.  Although we do not consider spin in this work, we attempt to leverage recent developments in quantum information theory to observe topological features predicted by the Laughlin and MR theory in numerically obtained FQHE ground states and examine how these features change under a realistic change (in particular, the finite thickness effect with varying thickness) in the effective interaction. 

These developments in quantum information theory provide the context for our current work.  In this article we provide a detailed numerical theoretical study of entanglement for FQHE states incorporating the dependence on the quasi-2D layer thickness of the transverse dimension (i.e the finite layer thickness effect\cite{MacDonald84, Zhang86, SongHe90, Ortalano97, Park98, PetersonPRL08,Peterson08}).
We emphasize physics of the finite-thickness effect in this article partly because it provides a qualitative understanding of the FQHE in higher Landau levels as shown in Ref.~\onlinecite{PetersonPRL08,Peterson08}.  Indeed, the orbital Landau level (LL) dependence of the FQHE is not completely understood~\cite{Kumar}.  The theory behind the MR Pfaffian model state, for example, makes no distinction between the half-filled second orbital Landau level (SLL) (i.e., $\nu = 5/2$) and the half-filled lowest orbital Landau level (LLL) ($\nu = 1/2$), but no FQHE has so far been observed at $\nu = 1/2$ in monolayer systems.  Furthermore FQH states in the SLL are relatively rare and generally ``weaker'' (i.e., requiring lower temperatures and higher sample mobilities to experimentally observe due to their relatively small energy gaps on the order of 0.5 K or less) compared to FQH states in the LLL\cite{Willett87, Pan99, Eisen02, Xia04, Csathy05, Choi08, Pan08}.  Some light was shed on this phenomenon of fragility of the FQHE in the SLL in Ref.~\onlinecite{PetersonPRL08,Peterson08} where it is shown that the finite-thickness effect is qualitatively dependent on Landau level.  In particular, it is shown that a finite, non-zero, layer thickness, $d$, helps stabilize the FQHE in the SLL, whereas in the LLL, finite $d$ tends to ``weaken" the state.  Also, for the $\nu=5/2$ state, there appears to be an optimal thickness where the MR Pfaffian is ``strongest.''  Thus, there seems to be a close connection between finite thickness and the SLL FQHE, which we explore in this work by calculating the thickness dependent entanglement properties of the FQHE.  The results given in Ref.~\onlinecite{PetersonPRL08,Peterson08} are primarily based on calculations of the overlap between numerically obtained ground states and model FQHE wavefunctions~\cite{footnote} --- in particular, the Laughlin and MR Pfaffian model states.  Although the overlap is a powerful theoretical tool --- in fact, the wide acceptance of the Laughlin wavefunction as the appropriate description of the experimentally observed odd-denominator FQHE is arguably based on overlap results --- it is not always definitive and can be misleading in some cases.  An example of this involves the $\nu=2/5$ state that has been shown to have a large overlap with both the Jain composite fermion wavefunction as well as the so-called ``Gaffnian" wavefunction, even though these two states have different underlying topological order belonging to different universality classes~\cite{Simon07,Regnault09}.   Therefore, we seek alternative theoretical tools to probe the Landau level dependence of the FQHE through the finite-thickness effect, which should inevitably give insight into the nature of the enigmatic $\nu=5/2$ state.  In this article, we examine bipartite entanglement as an alternative measure (in particular the entanglement entropy and the entanglement spectrum) and study its dependence on finite layer thickness, $d$, in quasi-2D FQH states.  One reason for our study of the thickness-dependent FQHE entanglement is that the thickness parameter $d$ enables a continuous tuning of the Hamiltonian, lending to a continuous variation in the entanglement allowing a comprehensive systematic study.
 
Bipartite entanglement measures are tools designed to quantify the extent to which degrees of freedom are entangled in a bipartitioned system\cite{Plenio07}.  The most straight forward of these is the entanglement entropy (EE), defined as the the von-Neuman entropy
\begin{eqnarray}
S_E = \mathrm{Tr} [\rho_A \ln \rho_A]
\end{eqnarray}
of the reduced density matrix $\rho_{A/B} = \mathrm{Tr}_{B/A}[\left|\Psi\right\rangle\left\langle\Psi\right|]$ for state $\left|\Psi\right\rangle \in \mathcal{H} = \mathcal{H}_A\otimes \mathcal{H}_B$ in a Fock space $\mathcal{H}$ that has been partitioned into two parts.  The EE has proven to be a very powerful tool in examining quantum correlations in interacting many-body systems\cite{Amico08}.  In particular, the scaling of the entanglement entropy with system size has been shown to follow certain ``area'' laws that can identify quantum phase transitions in certain cases\cite{Eisert10}.  Also, the EE can be used to extract the ``topological entanglement entropy'' which is an indicator of topological order in the system\cite{Kitaev06, Levin06} (i.e. states in the same topological class will have the same topological entanglement entropy).  Given these developments, the EE and the topological EE appear to be attractive tools to probe FQH states.  However, as we discuss below in more detail, obtaining a precise quantitative estimate of the topological entropy in FQH states requires a technical procedure prone to introducing significant errors~\cite{Zozulya07, Iblisdir, Lauchli2010}.  In our study, we do not attempt to make such an estimate.  Instead we focus on FQHE states with the primary objective of observing the qualitative behavior of the EE as a function of finite layer thickness, $d$, in the LLL and SLL.  Our goal is to obtain the qualitative dependence of the EE in FQHE as a function of layer thickness $d$.

We also investigate the ``entanglement spectrum'' of quasi-2D FQH states in this study.  Introduced by Li and Haldane\cite{LiHaldane}, the entanglement spectrum (ES) provides more information than the entanglement entropy alone.  In particular, Li and Haldane conjectured that there is a direct correspondence between the low-lying eigenvalues of the operator $\widehat{h}=-\ln[\rho_A]$ and the edge modes of the system and, thus, ES can be used to determine characteristics of the underlying conformal field theory (CFT) of the corresponding FQH ground state.  As long as these ``CFT-like'' states are well-separated from the ``generic non-CFT-like'' states in the entanglement spectrum of the ground state in the thermodynamic limit, then it is conjectured that the identified CFT does, indeed, describe the state.  In other words, states described the by the same CFT (i.e. in the same universality class) will have the same low-lying structure in their respective entanglement spectra. Numerical studies~\cite{Hermanns10,Chandran11, Zhao, Sterdyniak, PapicPRL} and some recent analytical results~\cite{Qi2011} support the Li and Haldane conjecture. In our study, we carefully examine the entanglement spectrum  of finite sized FQHE states by quantifying the separations between the suspected low-lying CFT and non-CFT-like states (``entanglement gaps'') as a function of finite-layer thickness.  The entanglement gaps serve as a semi-quantitative measure of how well the state in a finite system fits with the universality class described by the CFT. By doing this we discover that the entanglement gaps follow trends qualitatively similar to the EE, leading to similar conclusions obtained in the overlap calculations given in Ref.~\onlinecite{PetersonPRL08,Peterson08}  (i.e., that the finite-thickness effect strengthens the FQH states in the SLL).  We also examine the entanglement spectrum in the so-called ``conformal limit"~\cite{Thomale10}.  The aim of the conformal limit is to remove finite size artifacts from the geometry of the system (in our case, the sphere), allowing for a concrete definition of a full entanglement gap.  We find that the conformal limit does result in a full entanglement gap in most cases studied, but not in all situations.   The lack of a entanglement gap in these exceptional cases, however, is likely due to our choice of planar pseudo-potentials in our work\cite{Papic}; further work along this line would be necessary to fully understand these situations. 
  
The structure of this article is as follows: In section \ref{sec:Method} we describe our methods for numerically obtaining the exact ground state of the Coulombic Hamiltonian for FQH ground states at filling fractions $\nu = 1/3$, 7/3, 1/2, and 5/2 and our chosen model of the finite thickness of the quasi-2D system as well as the model wavefunctions (Laughlin and MR Pfaffian) to which we compare the numerically obtained ground state.  Further, we describe how our entanglement measures are defined and calculated.
In section \ref{sec:Results} we provide our results for the entanglement entropy (\ref{sec:EE}), entanglement gaps in the entanglement spectrum (\ref{sec:ES}), entanglement spectra in the conformal limit (\ref{sec:CL}) and density of states calculations of entanglement spectra (\ref{sec:DOS}) for FQH ground states at the filling fractions listed above.  We provide our conclusions in section \ref{sec:Conclusion}.   Finally in the appendix, we discuss our choice of planar Haldane pseudopotentials over spherical pseudopotentials and examine the implications of this choice by comparing entanglement spectra in FQHE ground states at zero thickness obtained with either choice of pseudopotentials.

\section{Method} \label{sec:Method}
We begin by considering a quasi-2D geometry where spinless electrons are confined in the x-y plane with layer thickness $d$ along the z-axis and an external magnetic field $B$ also along the z-axis.  In the non-interacting case, the presence of the magnetic field quantizes the electron energy levels into highly degenerate Landau levels (LL) with energies $E_n = (n+1/2)\hbar\omega_c$ where $\omega_c=eB/mc$ and $n=0,1,2,\ldots$ is a non-negative integer that defines the LL index.  The degeneracy of each LL per unit area is given by $(2\pi l^2)^{-1}$, where $l = \sqrt{\hbar c /eB}$ is the magnetic length and defines a length scale for the problem.  The filling factor  is given by $\nu = N_s/(2\pi l^2)$ where $N_s$ is the particle density per unit area.  When we include spin, the degeneracy of each Landau level is doubled.  If we assume the electrons are polarized by the magnetic field and the LL's are filled sequentially by spin up and spin down electrons, then $0 < \nu < 2$ corresponds to states in the lowest Landau level (LLL) with LL index $n=0$ and $2 < \nu < 4$ corresponds to the second Landau level (SLL) with LL index $n=1$.  Thus, the 1/2-filled LLL corresponds to $\nu=1/2$ or 3/2 and the 1/2-filled SLL corresponds to $\nu=5/2$ or 
7/2.  The identifications of the 1/3-filled filling factors is obtained similarly.  

The presence of electron-electron interactions clearly complicates this picture, but if we assume the electrons are confined to a single LL (i.e., there is no LL mixing), then the kinetic energy is a constant that can be ignored, giving the effective Hamiltonian:
\begin{equation}
\hat H = \sum_{i<j}^{N}V(r_{ij}), \label{eg:columbH}
\end{equation}
where $N$ is the total number of particles and $r_{ij} = |r_i - r_j|$ is the distance between particles measured in units of the magnetic length $l$.  Note that our assumption of no LL mixing might not be a very good approximation~\cite{Bishara,Simon-Rezayi,Toke-Jain} in all cases especially when considering the FQH state at $\nu=5/2$ where it has been observed at ``low" magnetic fields where the LL mixing parameter $[(e^2/(\epsilon l))]/\hbar\omega_c\sim 1$ ($e^2/\epsilon l$ characterizes the strength of the Coulomb interaction where $e$ is the electron charge and $\epsilon$ is the dielectric constant of the host semiconductor).  Furthermore, the Hamiltonian we are using is completely particle-hole symmetric and apparently the ground state does not spontaneously break particle-hole symmetry~\cite{PetersonParkDasSarma}.  This means that all of our conclusions about the MR Pfaffian state apply equally to the anti-Pfaffian state since the two ansatz are particle-hole conjugates.  LL mixing, as shown recently~\cite{Bishara,Simon-Rezayi,Toke-Jain}, can induce three-body terms which explicitly break particle-hole symmetry leading to a possible preference toward the MR Pfaffian or anti-Pfaffian, however, in this work we do not consider such terms and can make no distinction.

This Hamiltonian (Eq. (\ref{eg:columbH})) can be parametrized in terms of the relative angular momentum between two particles $m_{ij}$ by the Haldane ``pseudopotential'' expansion\cite{Haldane83}:
\begin{equation}
\hat H =  \sum_{i<j}^{N}V(r_{ij}) = \sum_{m(odd)} V_m^{(n)} \sum_{i<j}^{N}\hat P_m(m_{ij}) \label{eq:pseudoH}
\end{equation}
where $\hat P_m(m_{ij})$ projects onto states with relative angular momentum $m_{ij}=m$ and $V_m^{(n)}$ are the Haldane pseudopotentials for a given relative angular momentum $m$ and Landau level index $n$.  These pseudopotentials $V^{(n)}_m$ are the energies of a pair of particles with relative angular momentum $m$ confined to the $n$-th LL.   For spinless fermions, only $V_m^{(n)}$ for odd $m$ enter the Hamiltonian due to the Pauli exclusion principle.  In a planar geometry, the Haldane pseudopotentials are given in terms of the Fourier transform of the interaction potential $V(k)$ by
\begin{equation}
V_m^{(n)} = \int^{\infty}_{0}dk k [L_n(k^2/2)]^2L_m(k^2)e^{-k^2}V(k) \label{eq:pseudo}
\end{equation}
where $L_n(x)$ are Laguerre polynomials.   This expansion allows us to work strictly in the Hilbert space of the lowest Landau level since all necessary information about electrons confined in higher Landau levels are contained in the pseudopotentials.

The model wavefunctions (Laughlin and MR Pfaffian states) can be obtained by diagonalizing certain ``hardcore'' Hamiltonians.  The Laughlin wavefunction
\begin{eqnarray}
\Psi_{\mathrm{Laughlin}}=\prod_{i<j}^{N}(z_i-z_j)^q e^{-\sum |z_i|^2/4}
\end{eqnarray}
where $z=x-iy$ is the electron coordinate in the complex plane, and is the zero-energy ground state of a special case of Eq. (\ref{eq:pseudoH}). 
 For filling fraction $\nu = 1/q$, $q$ odd, this Hamiltonian is given by:
\begin{equation}
\hat H^{(q)}_L = \sum^{q-2}_{m(odd)}  \sum_{i<j}^{N}\hat P_m(m_{ij}). \label{eq:laughH}
\end{equation}
This two-body ``hardcore'' potential penalizes any state where two particles have a relative angular momentum smaller than $q$.  The MR Pfaffian wavefunction
\begin{eqnarray}
\Psi_{\mathrm{MR}}=\mathrm{Pf}\left\{\frac{1}{z_i-z_j}\right\}\prod_{i<j}^N(z_i-z_j)^2 e^{-\sum |z_i|^2/4}
\end{eqnarray}
 is the exact zero-energy ground state of a three body Hamiltonian projecting onto electron triplets instead of 
pairs~\cite{GreiterWenWilczek}.

In an ideal, strictly two dimensional system, the electron-electron interaction is given by the 2D Coulomb potential, $V(k)=(e^2l/\epsilon)(1/k)$.  The finite extent of an experimental quantum Hall system in the perpendicular direction will alter the ideal 2D interaction, yielding an effective quasi-2D electron-electron interaction.  There are several models for the effect of the finite layer thickness\cite{Tsuneya82, Stern84, DasSarma85, Zhang86}, however, these models all provide similar qualitative 
results~\cite{PetersonPRL08,Peterson08}.  Therefore, we will focus on one particular model, the infinite square well potential.  In this model, we average the three dimensional Coulomb potential over the single-particle ground state $n(z)$ of an infinite square well (i.e., $n(z) = \sqrt{2/d}\cos(\pi z/d)$) in the perpendicular dimension, yielding the effective interaction potential
\begin{eqnarray}
V_{SQ}(k) &=& \frac{e^2}{\epsilon l}\frac{1}{k}\int dz_1 dz_2 |n(z_1)|^2|n(z_2)|^2e^{-k|z_1-z_2|} \nonumber \\
&=&\frac{e^2l}{\epsilon k}\frac{\left\{3kd + \frac{8\pi^2}{kd}- \frac{32\pi^4(1-e^{-kd})} {(kd)^2\left[(kd)^2+4\pi^2\right]}\right\}}{(kd)^2+4\pi^2}. \label{eq:Veff}
\end{eqnarray}
Combining Eq. (\ref{eq:Veff}) with Eq. (\ref{eq:pseudo}) gives us effective pseudopotentials as a function of finite thickness $d$, Landau level $n$, and relative angular momentum $m$.

We diagonalize the FQH Hamiltonians (one for each $d/l$ and LL index $n$) in the spherical geometry\cite{Haldane83} where $N$ electrons are confined to the surface of a sphere.  Although we use this geometry, we use the pseudopotentials obtained from the infinite planar geometry (Eq.(\ref{eq:pseudo})) since the finite layer thickness effect is more conveniently modeled in this case.  Furthermore, the pseudopotentials in the spherical geometry equal those in the planar geometry as the thermodynamic limit is approached (as the spherical radius is taken to infinity) and it can be argued that they provide a better approximation to the thermodynamic limit  (this is discussed in detail in Ref.~\onlinecite{Peterson08} and in the appendix). In the spherical geometry, the perpendicular (radial) magnetic field is provided by a magnetic monopole of strength, $Q$, quantized in half-integer units, placed at the center of the sphere.  The eigenvalues of the squared magnitude $L^2$ and z-component $L_z$ of the angular momentum, $S(S+1)$ and $m$ respectively, are good quantum numbers for the single particle wavefunctions, where $S$ is related to the LL index by the constraint, $S = |Q| + n$ and $m$ is constrained such that $-S \leq m \leq S$.  Thus the degeneracy of a LL with index $n$ is given by $g = 2S + 1 = 2(|Q| + n) + 1$.  The filling factor for a LL is defined in the thermodynamic limit $\nu = \lim_{N\rightarrow\infty}N/g$.  The uniform ground state has total angular momentum, $L = 0$ and therefore, can be obtained in the Hilbert subspace where the total $z$-component of angular  momentum, $L_z=0$.   
 
In this study, we consider the FQH ground states at the Laughlin filling fractions $\nu = 1/3$ and $\nu = 7/3$ with particle number $N = 6$, 7, and 8 and the even-denominator filling fractions $\nu = 1/2$ and $\nu = 5/2$ with particle number $N = 8$ and $N = 10$.  We restrict ourselves to these relatively modest system sizes in order to investigate a large number of FQH ground states for various values of the finite thickness $d/l$ with reasonable computing resources.  Although the Hilbert space for particle number $N=12$ at the half fillings is not prohibitively large, this system is also aliased with $\nu = 2/3$ and could, therefore, 
yield ambiguous results. Since we are largely concerned with the qualitative features of the finite-thickness effect, these system sizes are adequate.     

\begin{figure}
\includegraphics[width=.47\textwidth]{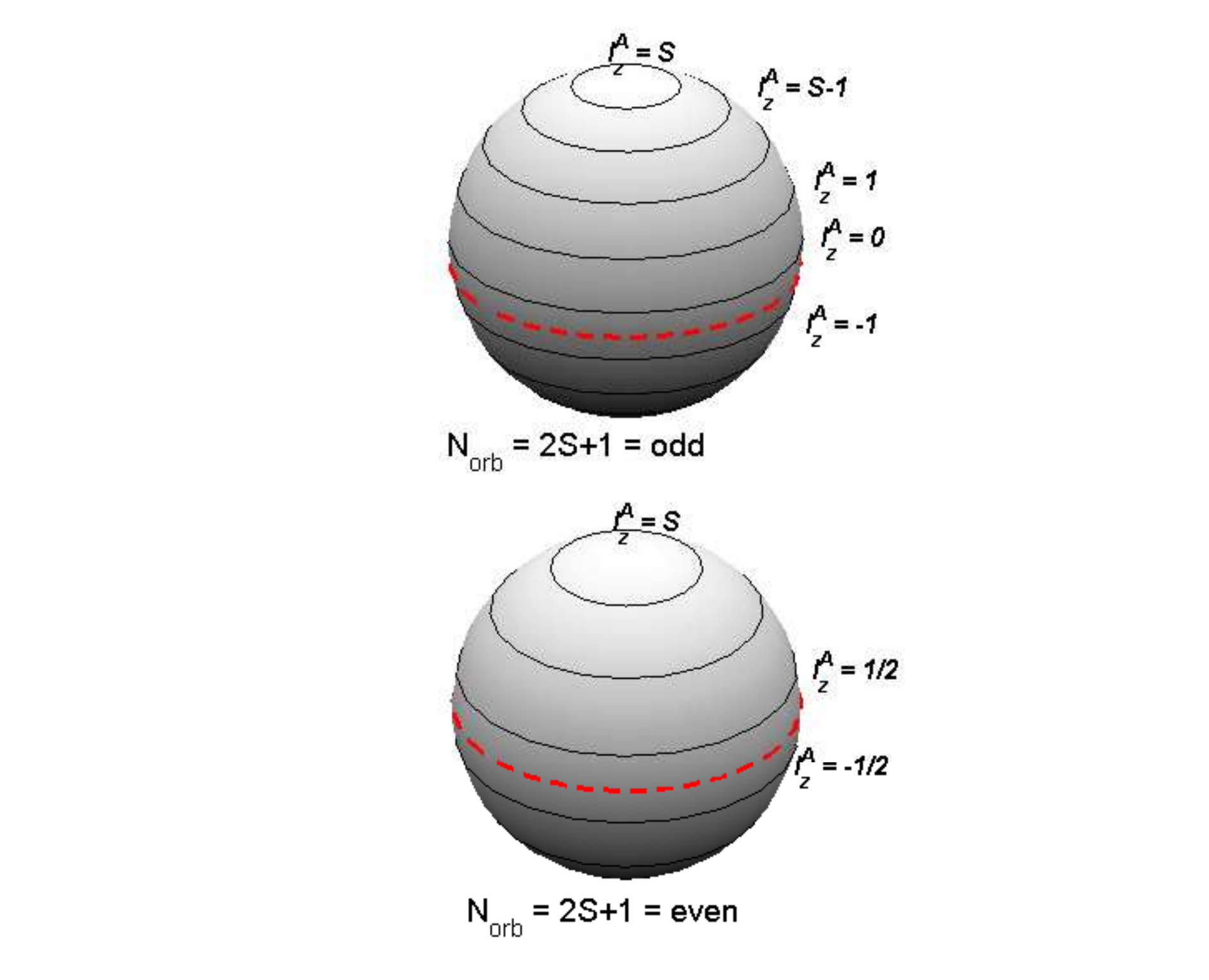}
\caption{\label{fig:sphericalcut} Graphical illustration of the partitioning of the Fock space.  In the spherical geometry, the single-particle states are states with the $z$-component 
of angular momentum  from $S$ to $-S$, represented by the solid latitudinal lines.  We choose our partitions to cut the sphere in two as close to the equator 
as possible, represented by the dashed lines.  Thus, for $N_\mathrm{orb}=2S+1$ even, we cut the sphere after $N_\mathrm{orb}/2$ (see top panel) and after 
$(N_\mathrm{orb}+1)/2$ for $N_\mathrm{orb}$ odd (see bottom panel).}
\end{figure}

We calculate the entanglement entropy (EE) and the entanglement spectrum (ES) of FQH ground states by dividing the sphere into two regions.  In particular we write our Fock space as the tensor product of two subspaces $\mathcal{H} = \mathcal{H}_A \otimes \mathcal{H}_B$ with respective basis states $\left|\psi_A^i\right\rangle = \left|n_{-S}^i, n_{-S+1}^i ... n_{-S+N_\mathrm{orb}^A-1}^i\right\rangle \in \mathcal{H}_A$ and $\left|\psi_B^k\right\rangle = \left|n_{S-N_\mathrm{orb}^B}^k, n_{S-N_\mathrm{orb}^B+1}^k ... n_{S}^k\right\rangle \in \mathcal{H}_B$ where $n_m$ is the occupation number of the Landau orbital with angular momentum, $m$ and $N_\mathrm{orb}^A+N_\mathrm{orb}^B=N_\mathrm{orb} \equiv 2S+1$ is the total number of Landau level orbitals.  For all cases we choose our partitions when dividing our Fock space $\mathcal{H}$ into $\mathcal{H}_A$ and $\mathcal{H}_B$ such that 
for the number of single particle orbitals $N_\mathrm{orb}=2S+1$ even,  $N_\mathrm{orb}^A=N_\mathrm{orb}^B=N_\mathrm{orb}/2$ and for $N_\mathrm{orb}$ odd, $N_\mathrm{orb}^A=N_\mathrm{orb}^B+1=(N_\mathrm{orb}+1)/2$.  Geometrically this is equivalent to dividing the sphere along a line of latitude (see Fig.~\ref{fig:sphericalcut}).

We calculate three quantities of interest: 

(i) The bipartite entanglement entropy EE of the ground state $\left|\Psi\right\rangle$ is given by $S_E = \mathrm{Tr}[\rho_A \log(\rho_A)]$ where $\rho_A$ is the reduced density matrix defined by $\rho_A = \mathrm{Tr}_B[\left|\Psi\right\rangle \left\langle\Psi\right|]$.  

(ii) The entanglement spectrum $\xi_i$ is obtained from the eigenvalues of the reduced density matrix, $\rho_i$, by the simple relation $\xi_i = 2\ln(|\rho_i|)$.  This is equivalent to finding the Schmidt-decomposition of the matrix $W_{ij}$ where $\left|\Psi\right\rangle = \sum_{i,j}W_{ij}\left|\psi_A^i\right\rangle \otimes \left|\psi_B^j\right\rangle = \sum_{i}\exp(-\xi_i/2)\left|\phi_A^i\right\rangle \otimes \left|\phi_A^i\right\rangle$, $\left|\phi_A^i\right\rangle \in \mathcal{H}_A$ and $\left|\phi_B^i\right\rangle\in \mathcal{H}_B$.  Since the quantum numbers for angular momentum and particle number in each region, $L_z^A, L_z^B, N^A$, $N^B$, are constrained such that $L_z^A+L_z^B=L_z=0$ and $N^A+N^B=N$, the reduced density matrix is block diagonal with $L_z^A$ and $N^A$ being good quantum numbers for the eigenstates of $\rho_A$.  Therefore $L_z^A$ and $N^A$ are good labels for the corresponding entanglement spectrum.

(iii) The  ``conformal limit" of the entanglement spectrum.  Recently, Thomale \emph{et al.}~\cite{Thomale10} have introduced the conformal limit when calculating the entanglement spectrum of spherical FQH ground states.  This limit is obtained by expressing the ground state in terms of a special choice of unnormalized basis states.  The normalized single particle wavefunction on a sphere with angular momentum $m$ is given by\cite{Haldane83,Fano86}
\begin{eqnarray}
\Psi(u,v)= \left(\frac{2S+l}{4\pi}{2S \choose S+m}\right)^{1/2}u^{S+m}v^{S-m},\nonumber 
\label{eq:singlewave}
\end{eqnarray}
where $u = \cos(\theta/2)e^{i\phi/2}$ and  $v = \sin(\theta/2)e^{-i\phi/2}$ with $\theta$ and $\phi$ the usual 
spherical coordinates.  In the conformal limit, we ``unnormalize" the single particle wavefunctions by removing the prefactor in Eq. (\ref{eq:singlewave}) such that the wavefunctions take the simple form $\Psi'(u,v) = u^{S+m}v^{S-m}$.  This procedure is an attempt to remove the finite size effects inherent in these calculations by basically removing the ``length" in the problem.  With the ground state redefined in this new basis, the entanglement spectrum is calculated as described above.

\section{Results} \label{sec:Results}
\subsection{Entanglement Entropy} \label{sec:EE}

We now report numerical results for the entanglement entropy (EE) of quasi-2D FQH ground states as a function of the finite layer thickness $d/l$ for FQH states in the LLL ($\nu=1/3$ and 1/2) and the SLL ($\nu=2+1/3=7/3$ and $2+1/2=5/2$).    As mentioned 
above we choose the partition to be as close to the equator of the sphere as possible to minimize finite size effects.  

\begin{figure*}
\includegraphics[width=.7\textwidth]{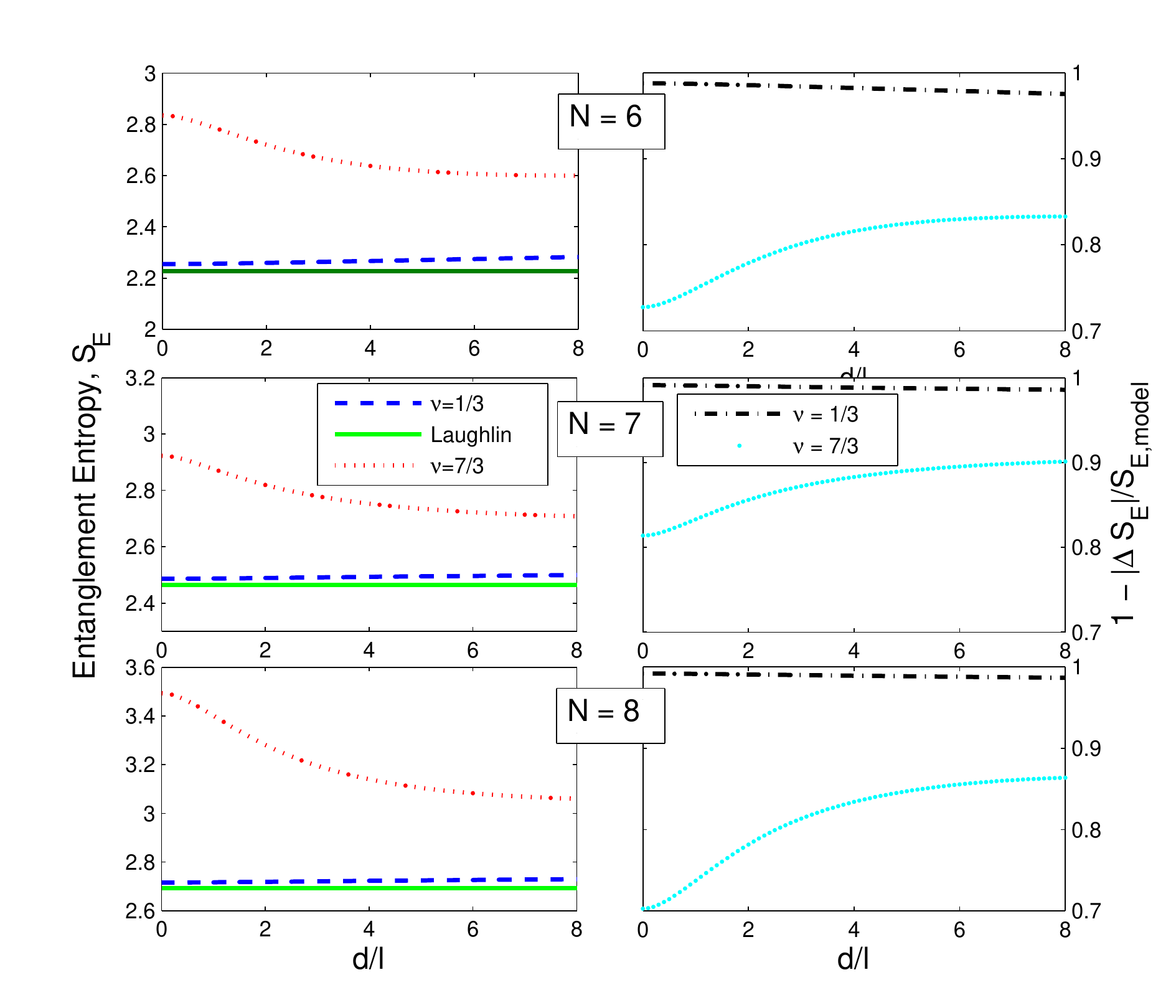}
\caption{\label{fig:EELaughlin3} (Color online) Entanglement entropy $S_E$ as a function of finite layer thickness, $d/l$ for Laughlin filling fractions $\nu=1/3$ and $\nu=7/3$ for particle number $N=6$, 7 and 8.  The dashed and dotted lines in the left panels correspond to the Coulomb Hamiltonian of a quasi-2D system in the LLL and SLL, respectively, whereas the solid lines in the left panels corresponds to the finite size Laughlin states.  The plots in the right panels give one minus the percentage difference in the Coulomb EE and the model state EE, i.e., $1-|\Delta S_E|/S_{E,\mathrm{model}}$ in the LLL (dash-dotted line) and the SLL (dotted line) and are found to be similar qualitatively and quantitatively to overlap calculations~\cite{MacDonald84,Zhang86,PetersonPRL08,Peterson08}.}
\end{figure*}

\begin{figure*}
\includegraphics[width=.7\textwidth]{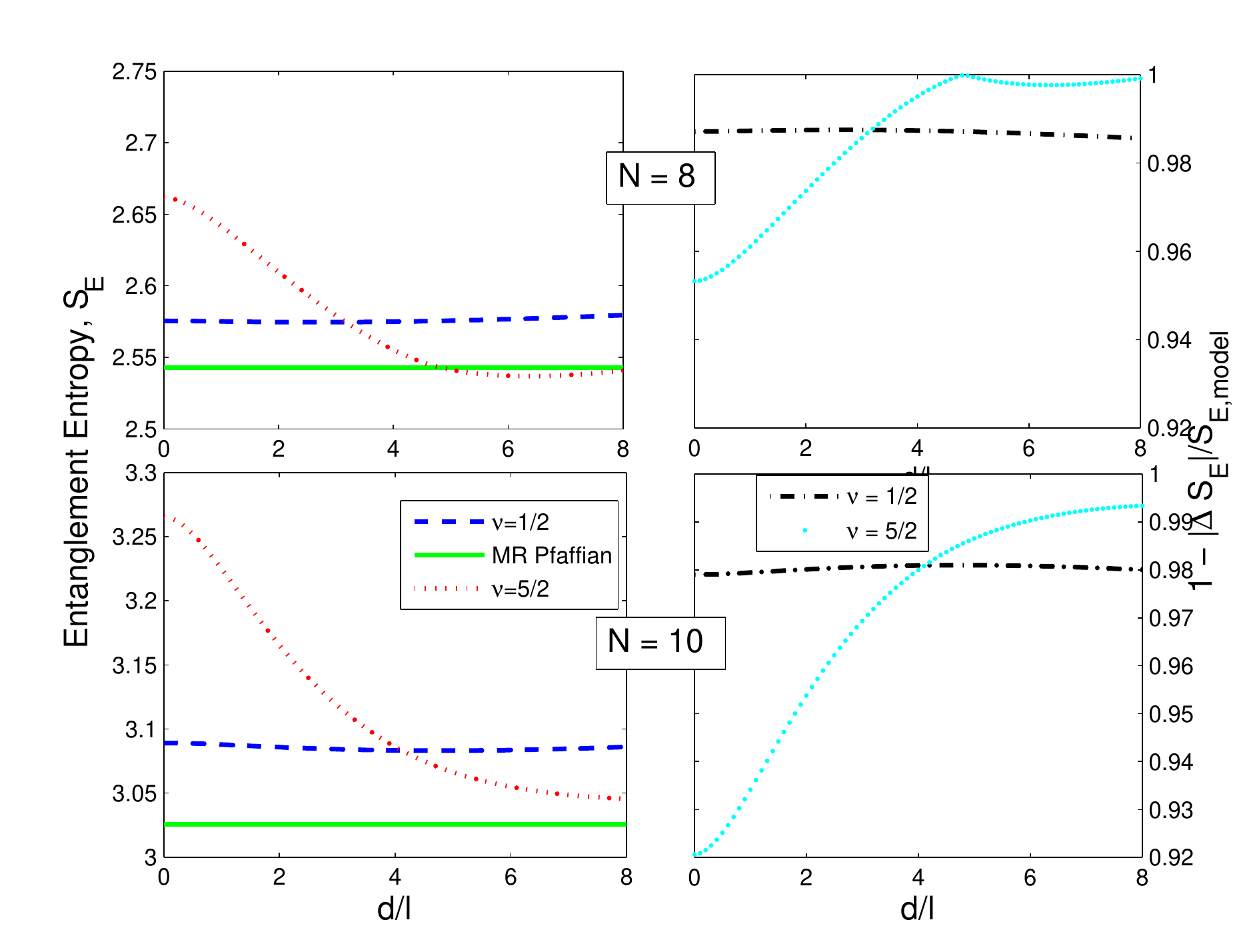} 
\caption{\label{fig:EEPfaffian} (Color online) Entanglement entropy $S_E$ as a function of finite layer thickness, $d/l$ for even-denominator filling fractions $\nu=1/2$ and $\nu=5/2$ for $N=8$ and 10.  Similarly to Fig.~\ref{fig:EELaughlin3} the plots in the right panels give one minus the percentage difference between $S_E$ and $S_{E,\mathrm{model}}$.  }
\end{figure*}

The results for EE for the Coulomb ground state at filling fractions $\nu= 1/3$ and $\nu=7/3$ are shown in 
Fig.~\ref{fig:EELaughlin3} as a function of finite layer thickness $d/l$.  For comparison, the EE of the corresponding Laughlin model wavefunction is also shown 
as a $d/l$ independent horizontal line.  In each of the figures, we see that in the LLL, the EE is near that of the Laughlin model wavefunction at $d=0$ and rises slightly as a function of $d/l$.  In contrast, the EE in the SLL is large compared to  that of the Laughlin model at $d=0$, but decreases as a function of $d/l$ and evidently reaches an asymptotic value.  The qualitative behavior is independent of system size.  If we consider $\Delta S_E = S_E - S_{E,\mathrm{model}}$ for both the LLL and SLL filling fractions and speculate that $\Delta S_E$ is a qualitative measure of how far removed the ground state is from the Laughlin model state, then we see that the LL dependence of $\Delta S_E$ as a function of $d/l$ behaves qualitatively similar to that of the overlap between the ground state and the model wavefunction as reported 
in Ref.~\onlinecite{MacDonald84,Zhang86,PetersonPRL08,Peterson08}.  In particular, the ground state in the LLL is a ``strong" FQHE state (i.e., $\Delta S_E$ is small) at $d=0$ and gradually becomes ``weaker" for increasing $d/l$ (albeit only slightly), whereas in the SLL, the ground state is initially weak at $d=0$ but gets stronger with increasing $d/l$ (i.e., $\Delta S_E$ decreases).   Thus, the EE for these cases 
qualitatively and semi-quantitatively captures 
how well the states are ``Laughlin-like" as a function of $d/l$ in similar manner to the overlap.  

Our operational definition of ``weak" and ``strong" depends on how close the EE of the Coulomb state is to the 
model state which, in this case, is the Laughlin state.  In the SLL, $S_E$ becomes closer to the $S_E$ for the Laughlin 
state but, as mentioned above, appears to saturate at some asymptotic value that is still nearly $\sim 1.1S_{E,\mathrm{Laughlin}}$.  
 In contrast, 
the EE in the LLL is almost identical to that of the Laughlin state.  We conjecture as to the reason 
for this difference between the EE in the SLL Coulomb ground state compared to the Laughlin state and the difference between the EE in the 
LLL as compared to the SLL:  (i) it is possible that the FQHE at 7/3 is \emph{not} described by the Laughlin 
state and is instead described by a state in a different topological universality class such as those given by Read and Rezayi~\cite{read-rezayi} and Bonderson and Slingerland~\cite{BS}, (ii) perhaps composite fermion interactions, which are thought~\cite{Toke-MRP} to be 
more relevant in higher LLs, are producing this difference in $S_E$ and the Laughlin state, (iii) perhaps the 7/3 FQHE 
state is in fact a Laughlin state but our model system is leaving out realistic effects such as LL mixing which 
are crucial to its success.

Fig.~\ref{fig:EEPfaffian} gives results for the EE of FQH ground states with even denominator filling fractions $\nu = 1/2$ and $5/2$ as a function of finite layer thickness $d/l$.  Also shown in the figure is the EE of the Moore-Read Pfaffian state for comparison.  In the LLL ($\nu = 1/2$), the EE has a weak minima as a function of $d/l$, in contrast to the Laughlin fractions (this minimum is difficult to discern on our scale).  The location of this minima changes with $N$, suggesting a finite size effect, but the qualitative behavior is similar in both cases.  In the SLL ($\nu = 5/2$), the EE has a very pronounced minima that approaches the EE of the MR Pfaffian model state for $N=10$ and crosses it for $N=8$.  This suggest that the FQH states becomes more MR Pfaffian-like at near an optimal $d/l$.  However, this optimal $d/l$ also changes with $N$.
Similarly to the Laughlin fractions, this LL dependence in EE as a function of $d/l$ is also qualitatively similar to that seen in the overlap between the FQH ground states and MR Pfaffian reported in Ref. \onlinecite{PetersonPRL08,Peterson08}.  These results suggest that $\nu = 1/2$ is not particularly well-described by the MR Pfaffian, whereas $\nu = 5/2$ is better described by the MR Pfaffian model state at finite thickness.

We note that recently, entanglement entropy in the SLL including finite thickness effects has been investigated~\cite{Friedman1,Friedman2}.   However, the previous study did calculations using the torus geometry, in contrast to our spherical geometry, and attempted to isolate and calculate the so-called topological term of the entanglement entropy.  The entanglement entropy can be essentially divided into two pieces
\begin{eqnarray}
S_E = \alpha L - \gamma + O\left(\frac{1}{L}\right)
\end{eqnarray}
where $L$ is the linear length of the boundary dividing the system into parts A and B (in our case it would be circumference of the sphere where we made our cut).  The $\alpha L$ term is non-topological in origin and the $-\gamma$ term is the topological entropy and for the Laughlin and MR Pfaffian state 
can be calculated analytically~\cite{Zozulya07}:  $\gamma=\ln{\sqrt{m}}$ for the Laughlin state and $\gamma=\ln{\sqrt{4m}}$ for the MR Pfaffian state at $\nu=1/m$.  Extracting the topological entropy from a Coulomb Hamiltonian requires one to  numerically calculate the exact ground state for many different systems sizes and system cuts and perform a thermodynamic extrapolation.  This is a labor intensive procedure that inherently induces statistical errors.  Such numerical extrapolation, without some strong theoretical guidance about the finite-size behavior of the system, is often unreliable in estimating quantities in the thermodynamic limit.

The conclusion of Refs.~\onlinecite{Friedman1,Friedman2} was that the topological entropy of the ground states of the LLL or SLL Coulomb Hamiltonians was consistent with associated model states (we note, however, that in Ref.~\onlinecite{Friedman1} it was concluded that $\nu=7/3$ was more consistent with the $k=4$ Read-Rezayi state~\cite{read-rezayi} instead of the Laughlin state).  However, they also included finite thickness in the form of an infinite square well potential and, interestingly, found that there was not much difference between the EE and the topological entropy with or without finite thickness included.  We, however, clearly see a finite thickness effect on the EE.  It is possible that this difference in the two studies (our present study and Refs.~\onlinecite{Friedman1,Friedman2}) is due to the different geometry used in the calculations (sphere vs. torus) but we find this scenario unlikely since most quantities of interest produce consistent results in the two geometries~\cite{haldane-rezayi,PetersonPRL08,Peterson08}.   Such a comparison between geometries (torus vs sphere) was shown in Ref.~\onlinecite{LauchliPRL2010} to give similar results for the entanglement spectra of Laughlin states, supporting our suspicion.  Moreover results given in Ref. ~\onlinecite{Lauchli2010} suggests that the the extrapolation procedure performed in Refs.~\onlinecite{Friedman1,Friedman2} may have been inappropriate for the torus. More work is clearly necessary to understand the difference between the results in spherical and toroidal geometries, particularly in the presence of the realistic finite thickness effects.

Before moving on to entanglement spectra we briefly discuss how our results compare to the previous overlap calculations done in Refs.~\onlinecite{PetersonPRL08,Peterson08}.   The right panels in Figs.~\ref{fig:EELaughlin3} and~\ref{fig:EEPfaffian} gives one minus the percentage error in the entanglement entropy, $1-|\Delta S_E|/S_{E,\mathrm{model}}$.   In Ref.~\onlinecite{Peterson08} it is found that the overlap between the Laughlin state and the Coulomb ground state at 1/3-filling in the LLL and SLL is approximately $\sim0.99$ at $d/l=0$ and is reduced monotonically to $\sim0.98$ at $d/l=8$ in the LLL and is $\sim0.73$ at $d/l=0$ and has a maximum of $\sim0.84$ for $d/l\sim 4$ in the SLL.  These overlap trends are very consistent with what we have seen previously in EE.  For the 1/2-filled LLL and SLL we find~\cite{PetersonPRL08,Peterson08} the overlap is relatively constant in the LLL at $\sim0.9$ and in the SLL it is $\sim0.96$ at $d/l=0$ and has a maximum value of nearly $\sim1$ at $d/l\sim 4$.  Again, one minus the percentage error in the entanglement entropy tracks the behavior in the overlap to a remarkable degree.  Perhaps this is not a surprise since if the overlap $\langle \Psi_0|\Psi_\mathrm{model}\rangle$ 
is close to one then the EE (which is a particular combination of $|\Psi\rangle\langle\Psi|$) should also be nearly identical to the EE of the model state $\Psi_\mathrm{model}$.

\subsection{Entanglement Spectrum} \label{sec:ES}
In the previous section, we saw that the entanglement entropy $S_E$ (and in particular, $\Delta S_E$) as a function of $d/l$ behaves
qualitatively similarly to the overlap~\cite{PetersonPRL08,Peterson08}.  For the half-filled case, increasing $d/l$ makes the 
calculated $S_E$ closer to the MR Pfaffian state for the SLL ($\nu=5/2$) in a rather dramatic way while increasing $d/l$ has 
very little effect on the $S_E$ in the LLL ($\nu=1/2$), i.e., using the entanglement entropy as a measure we see that the 
MR Pfaffian is \textit{stabilized} by finite thickness.  For the 1/3-filled case we find that increasing $d/l$ drives 
$S_E$ away from the Laughlin value in the LLL ($\nu=1/3$) and closer to the Laughlin value in the SLL ($\nu=7/3$), however, 
as in the previous overlap investigations, the value of the entanglement entropy for the $7/3$ case never gets as close to the 
Laughlin value as the 5/2 entanglement entropy gets to the MR Pfaffian.  As discussed above, this could be a hint 
that something is  missing from our understanding of the physics for the FQHE at $\nu=7/3$.

To gain a deeper understanding of entanglement, we now turn our attention to the finite layer thickness dependence of the entanglement spectrum (ES), which as discussed earlier, provides more information than the EE alone.  To calculate the ES, we partition the sphere the same as was done for the EE.   We follow the convention established by Li and Haldane\cite{LiHaldane} and restrict ourselves to the part of the ES where the number of particles in the $A$ partition, $N^A$, is the same as that of the ``root" configuration for the corresponding Laughlin or Moore-Read Pfaffian model wavefunction\cite{Bernevig08,LiHaldane} for a given partition size $N_\mathrm{orb}^A$.  The ``root" configurations describe the occupancy of LL orbitals for MR Pfaffian and Laughlin model states in the thermodynamic limit.  Root configurations with a maximum $z$-component of angular momentum, and their corresponding quantum numbers, $N^A$ and $L_z^A$, are given in Table \ref{tab:rootconfigs} for different filling fractions and partition sizes.
\begin{table}
\begin{tabular*}{.45\textwidth} {@{\extracolsep{\fill}} c | c | l | c | c}
\hline 
FQH state & $N_\mathrm{orb}^A$ & root config.  & $L_z^A$ & $N^A$ \\
\hline \hline
Laughlin 1/3,7/3   & 8           & `10010010'    & 13.5    & 3    \\ \cline{2-5}
      						 & 10          & `1001001001'  & 18      & 4    \\ \cline{2-5}
      						 & 11					& `10010010010' & 24      & 4    \\ \hline
MR Pfaffian 1/2, 5/2  & 7           & `1100110'     & 16      & 4    \\ \cline{2-5}
									 & 9           & `110011001'   & 24.5    & 5    \\
\hline  
\end{tabular*}
\caption{\label{tab:rootconfigs} Root configurations of the Laughlin ($\nu=1/3$, 7/3) and MR Pfaffian ($\nu=1/2$, 5/2) wavefunctions for the given partition sizes, $N_\mathrm{orb}^A$, on the sphere.  $`10010010'$, for example, means that the single-particle angular 
momentum $S=7.5$, $S-3=4.5$, and $S-6=1.5$ are all occupied with the others unoccupied.  Hence, there are $N^A=3$ electrons 
with total $z$-component of angular momentum $L_Z^A=3S-9=13.5$ in this root configuation.}
\end{table}
      
\begin{figure*}
\includegraphics[width=.65\textwidth]{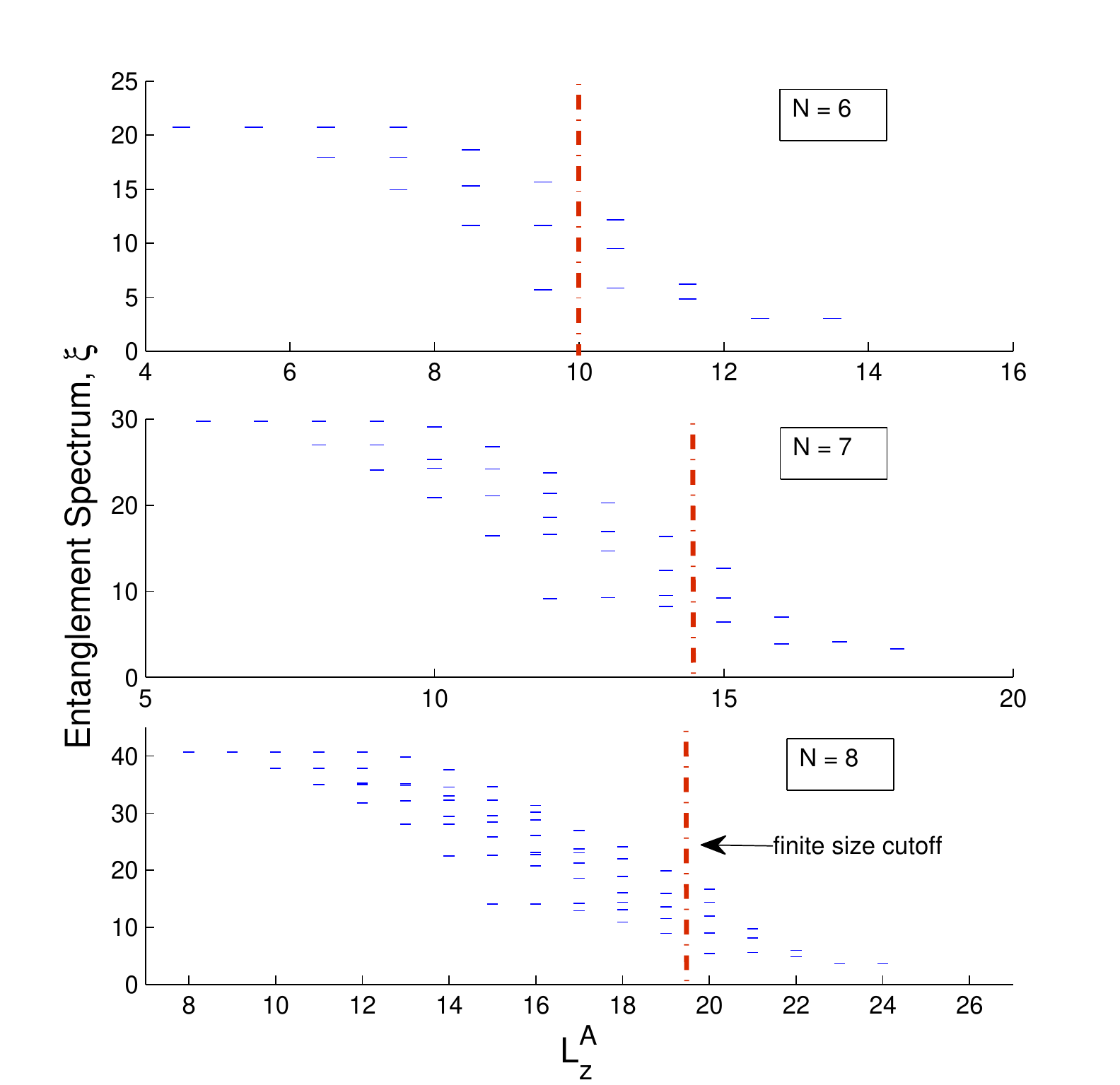}
\caption{\label{fig:ESLaughModel} Entanglement spectrum of the $\nu = 1/3$ filled Laughlin model state for particle number $N = $ 6 (top panel), 7 (middle panel), and 8 (bottom panel).  The finite size cutoff used to examine the entanglements gaps is illustrated by the vertical line.}
\end{figure*}

\begin{figure*}
\includegraphics[width=.65\textwidth]{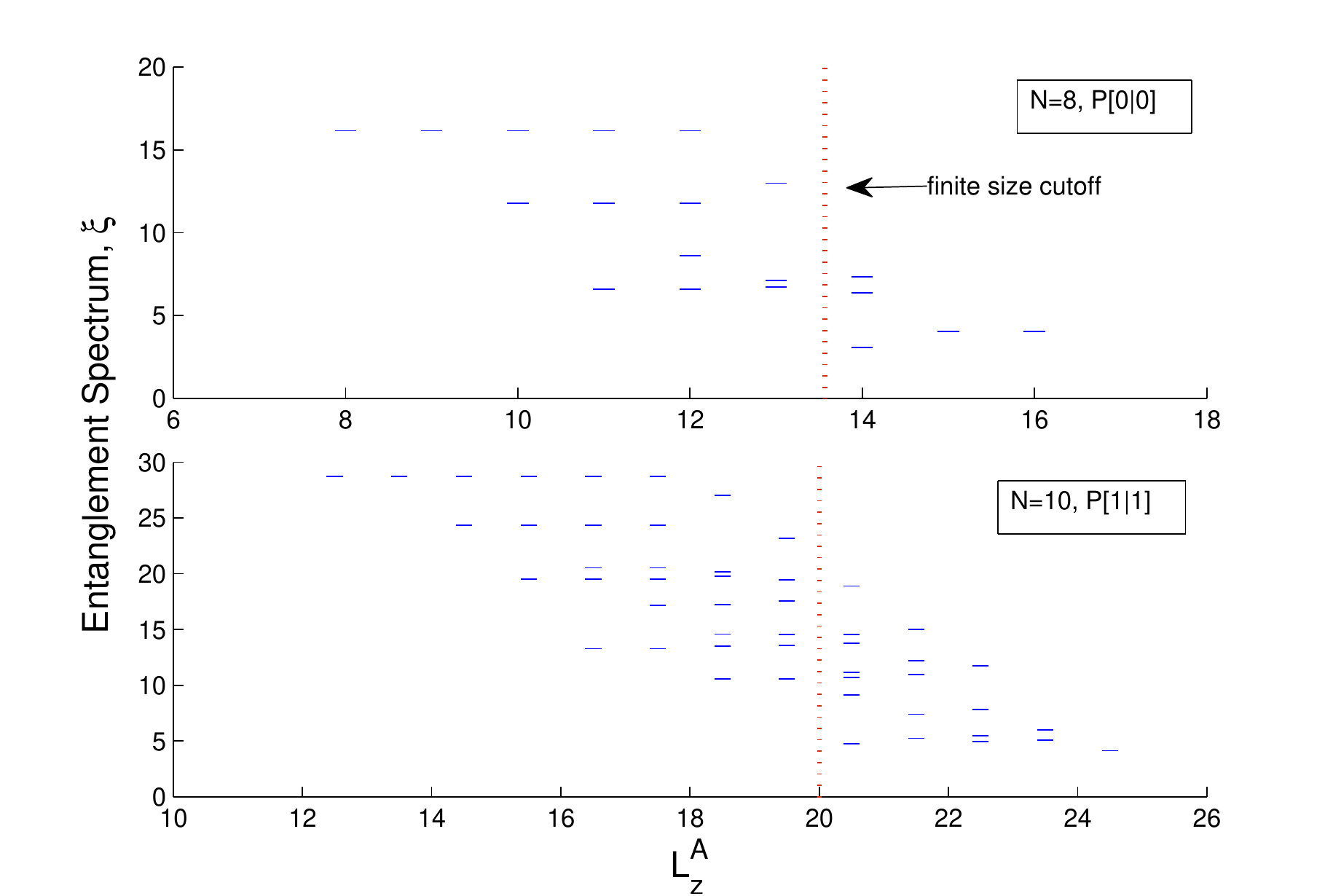}
\caption{\label{fig:ESmodelPfaff} Entanglement spectrum of the $\nu = 1/2$ filled MR Pfaffian model state for particle number $N = $ 8 (top panel) and 10 (middle panel).  The finite size cutoff used to examine the entanglements gaps is illustrated by the vertical line and the particular partition of the Fock space is given using the Li and Haldane notation~\cite{LiHaldane}.}
\end{figure*}

In order to obtain a general qualitative picture of how the ES changes as a function of the finite layer thickness, we calculate the ``entanglement gaps" in each ES and plot it as a function of $d/l$.  An entanglement gap\cite{LiHaldane} is defined as the difference between the low-lying ``CFT levels" (i.e., those levels displaying the low lying CFT counting structure) and the higher non-CFT levels for a given value of $L_z^A$ in the spectrum.  According to the Li and Haldane conjecture, the state has an underlying CFT if the entanglement gaps are finite in the thermodynamic limit.  However only the entanglement gaps at relatively ``small" values of $\Delta L = L_{z,\mathrm{root}}^A-L_z^A$ are relevant due to finite size effects, where $L_{z,\mathrm{root}}^A$ is the total $z$-component of angular  momentum of the root configuration.  The finite number of LL orbitals limits the number of possible ``edge excitations.''  Therefore only a few levels are expected to have the same counting structure as the CFT edge modes.  The ``depth'' (i.e., the max $\Delta L$) at which the counting structure in the ES is consistent with the CFT edge modes is dependent on the system size, $N$.  It has been conjectured~\cite{Hermanns10} that these ``finite size" effects of the entanglement spectra contain information such as the generalized statistics of the underlying FQH state, however, we will not consider such conjectures in this work.

We can determine a suitable cutoff for $\Delta L$ by examining when the level counting in the ES of the model states deviate from the expected counting in the thermodynamic limit.  To illustrate this finite size cutoff, we give the ES of the Laughlin state
in Fig.~\ref{fig:ESLaughModel}.  For the Laughlin state, the multiplicity of CFT levels is given by $\mathrm{p}(\Delta L)$ where $\mathrm{p}(m)$ is the partition function of the integer $m$.  The first 7 values of $\mathrm{p}(m)$, starting with $m=0$ are 1, 1, 2, 3, 5, 7, and 11.  In Fig.~\ref{fig:ESLaughModel}, we see that for $N = 6$ and 7, the level counting begins to deviate from $\mathrm{p}(\Delta L)$ at $\Delta L$ = 4, and for $N=8$, the deviation begins at $\Delta L = 5$.  Thus, for our study we will focus on the entanglement gaps for $\Delta L = 0, 1$, 2 and 3 for the $N = 6$ and 7 Laughlin systems, and for $N=8$, we also examine the entanglement gap at $\Delta L = 4$.

We determine the finite size cutoff for the entanglement gaps of the half-filled FQH states in a similar manner, which we now illustrate. The ES for the MR Pfaffian model states are shown in Fig.~\ref{fig:ESmodelPfaff}.  The counting rules for the MR Pfaffian model state depend on where the partition is made, which correspond to choosing one of the three sectors of the corresponding CFT\cite{LiHaldane}.  For the case of $N=8$, the partition along the equator is equivalent to the $P[0|0]$ partition in Li and Haldane's nomenclature (i.e., a cut between two unoccupied orbitals in the root configuration).  The CFT level counting for the first 4 Virasoro levels of this partition are 1, 1, 3, and 5.  The counting in the MR Pfaffian ES given in Fig.~\ref{fig:ESmodelPfaff} with $N=8$ deviates from this structure at $\Delta L = 3$.  For $N=10$, the partition along the equator corresponds to $P[1|1]$ (i.e., a cut between two occupied orbitals of the root configuration), which has a CFT level counting of 1, 2, 4, and 7 for the first 4 Virasoro levels.  Examining the ES in the figure for $N=10$, we see this spectrum also deviates from the expected counting at $\Delta L = 3$.  Thus for the half-filled FQH states we examine in this study, we concern ourselves only with the entanglement gaps up to $\Delta L = 2$.                      

\begin{figure*}
\includegraphics[width=.7\textwidth]{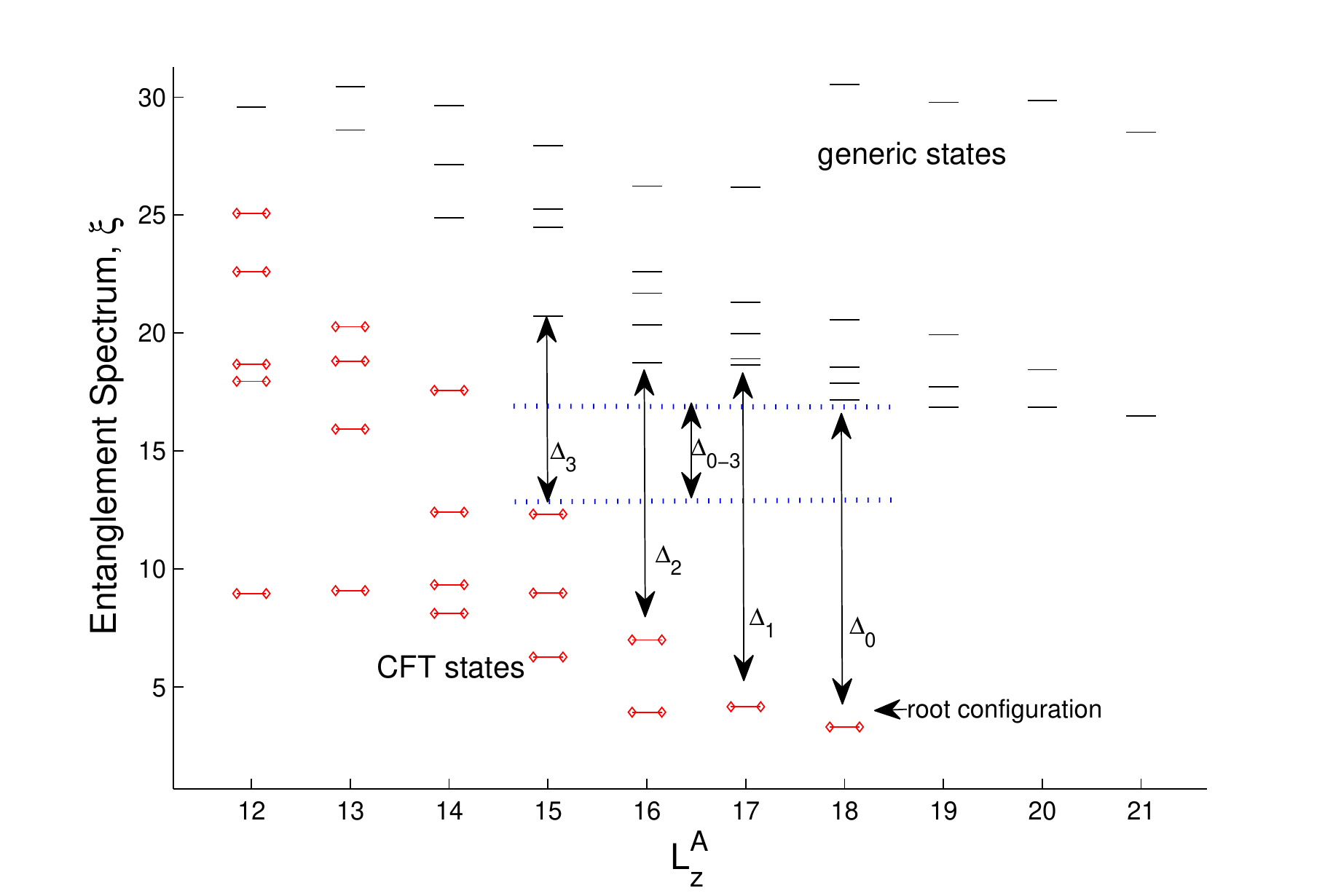}
\caption{\label{fig:ESexample} (Color online) Entanglement spectrum for $\nu = 1/3$, $d=0$, and $N=7$ as a function of $z$-component of angular momentum, $L_z^A$.  Suspected CFT and generic states are labeled by red (gray) diamonds connected by a red (gray) dash and black dashes, respectively, as well as the entanglement gaps $\Delta_n$.  The entanglement gaps are given by the difference between the lowest generic state and the highest CFT state for a given value of $L_z^A$.  The minimal gap, $\Delta_{0-4}$ is given by the difference between the lowest generic state and the highest CFT state for $\Delta L \leq 4$.  This minimal gap is represented by the dashed lines.  }
\end{figure*}

\begin{figure*}
\includegraphics[width=.65\textwidth]{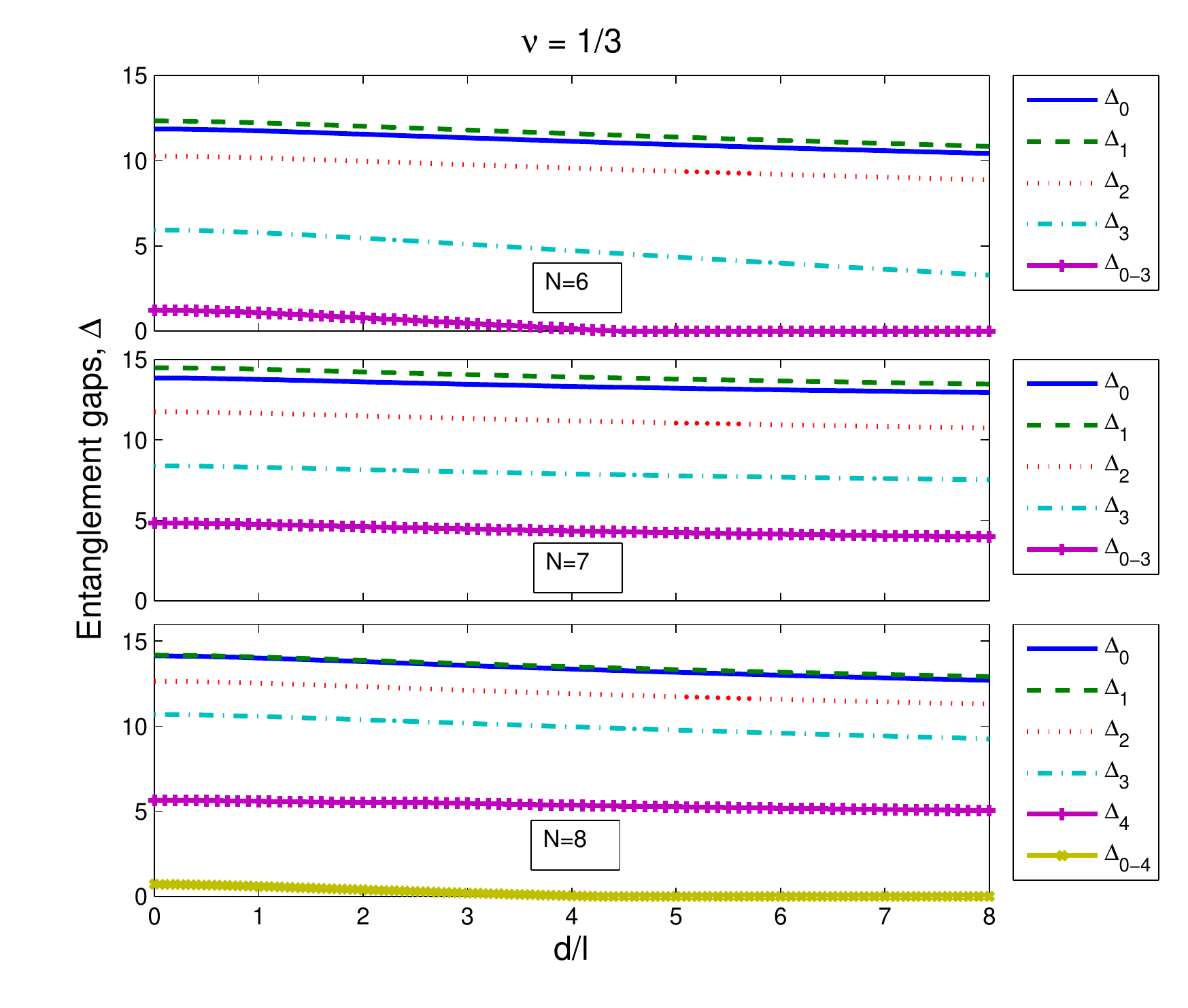}
\caption{\label{fig:ESGaps1_3} (Color online) Entanglement Gaps for the Coulomb Hamiltonian as a function of finite layer thickness $d/l$ for filling fraction $\nu = 1/3$ and particle number $N = 6$ (top panel), $N = 7$ (middle panel), and $N = 8$ (bottom panel) with partition at the equator. }
\end{figure*}

\begin{figure*}
\includegraphics[width=.9\textwidth]{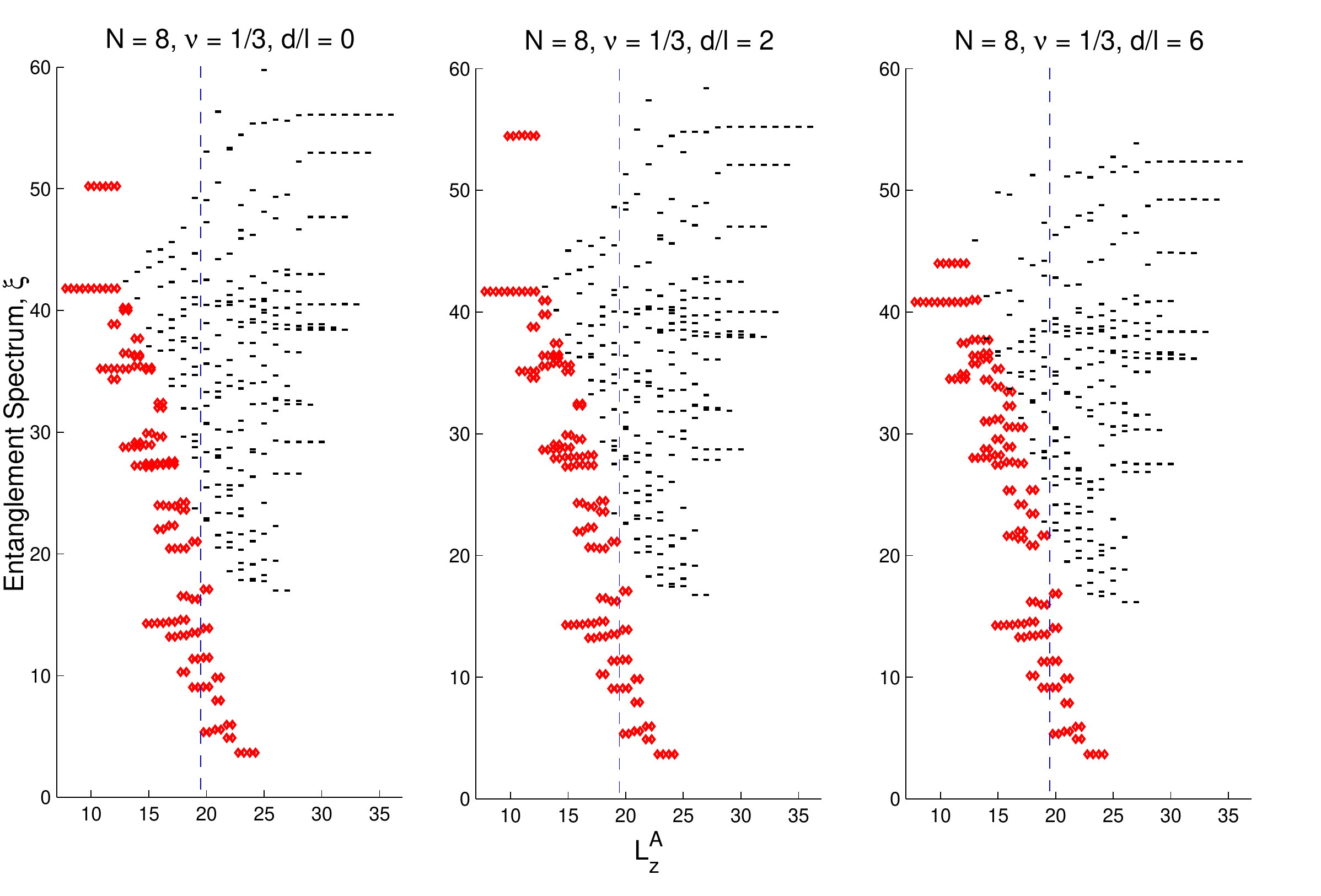}
\caption{\label{fig:ES1_3N8} (Color online) Entanglement spectrum for the Coulomb Hamiltonian as a function of $z$-component of angular  momentum $L_z^A$ for filling fraction $\nu = 1/3$ and particle number $N=8$ for $d/l$ = 0 (left panel), $d/l$ = 2 (middle panel), and $d/l$ = 6 (right panel).  The suspected CFT levels consistent with the Laughlin model state for each $L_z^A$ are marked by the diamonds connected by a dash.}
\end{figure*}

\begin{figure*}
\includegraphics[width=.65\textwidth]{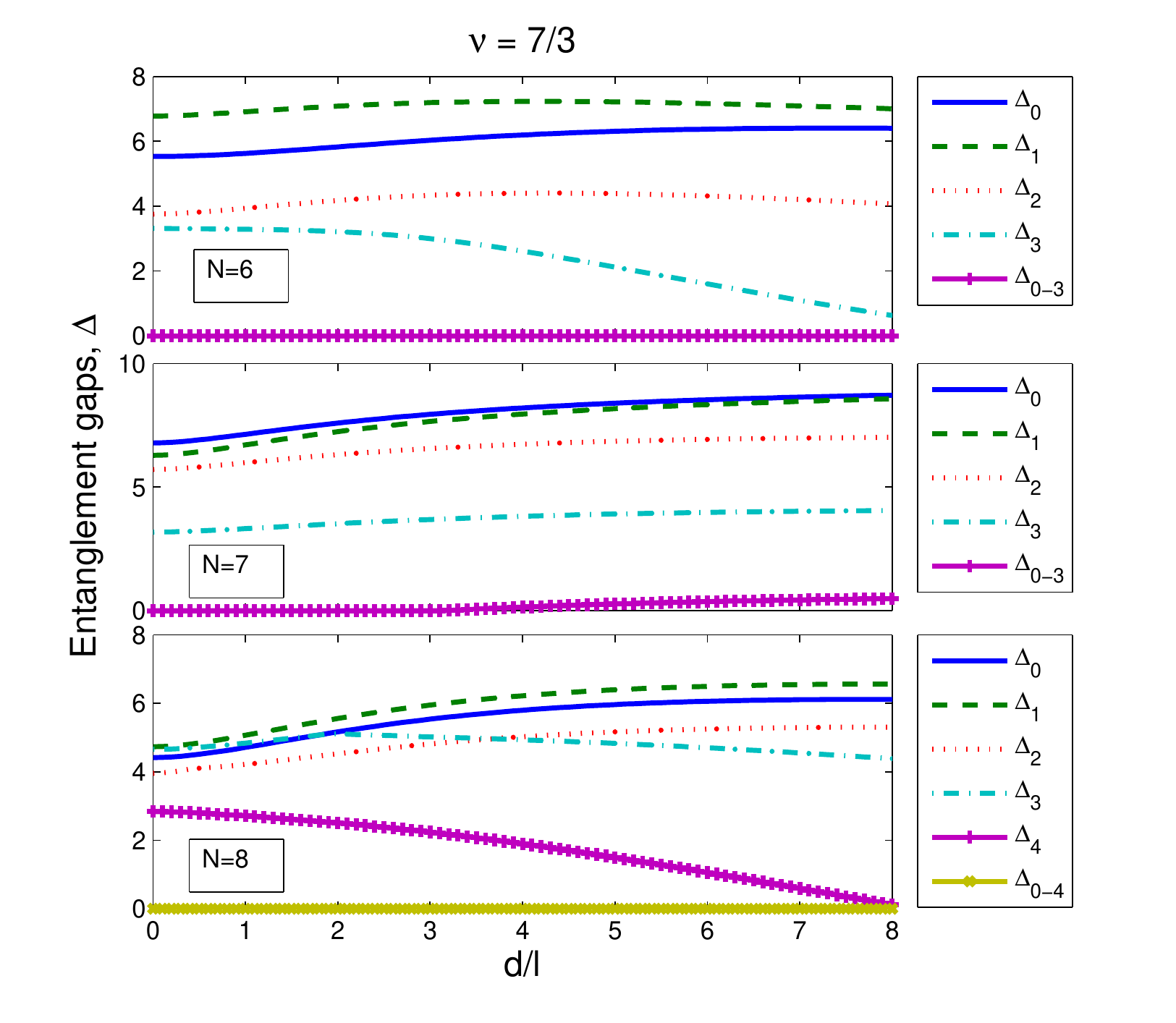}
\caption{\label{fig:ESGaps7_3} (Color online) Entanglement Gaps for the Coulomb Hamiltonian as a function of finite layer thickness, $d/l$ for filling fraction $\nu = 7/3$ and particle number $N = 6$ (top panel), $N = 7$ (middle panel), and $N = 8$ (bottom panel) with partition at the equator. }
\end{figure*}

\begin{figure*}
\includegraphics[width=.9\textwidth]{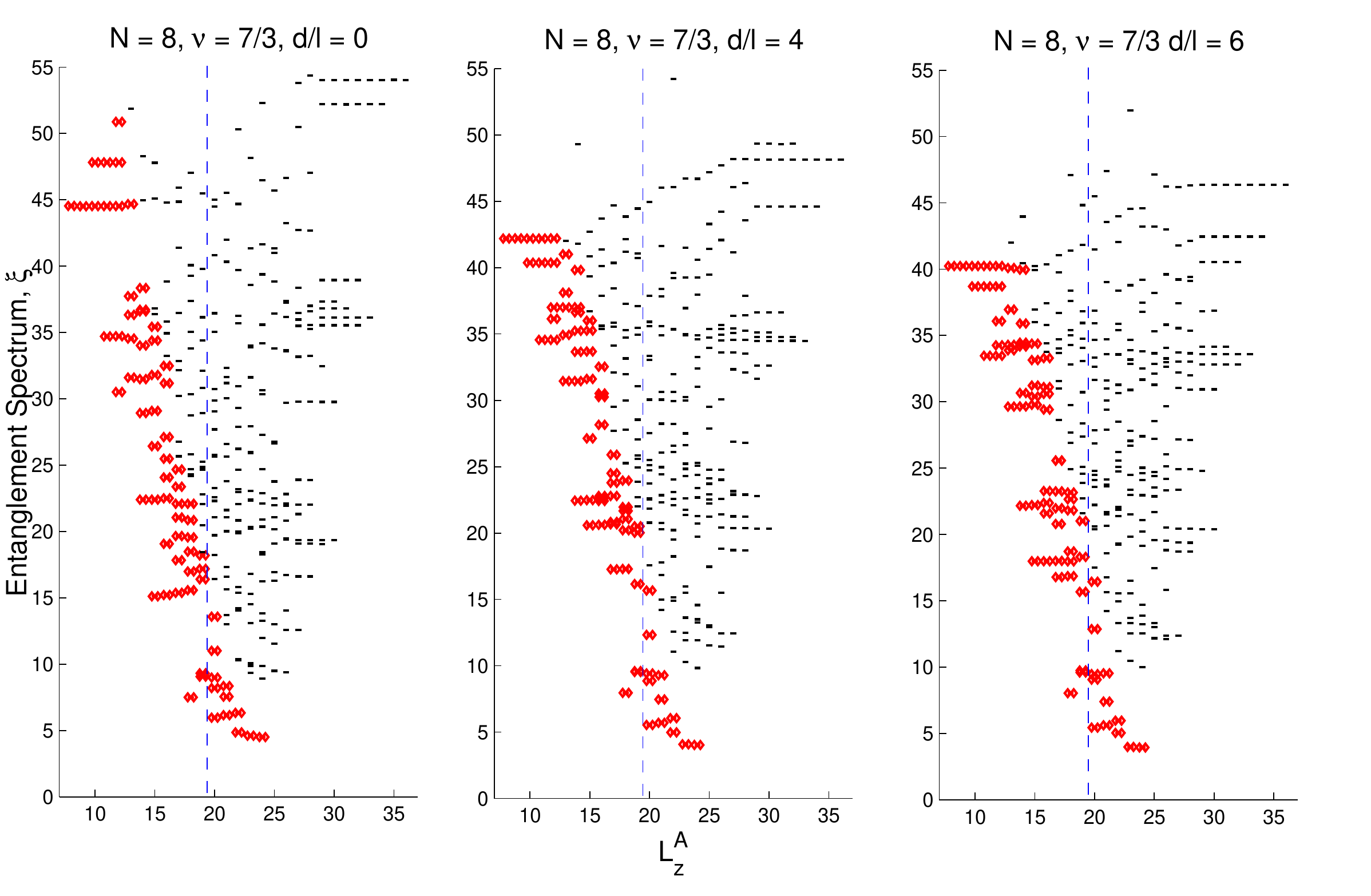}
\caption{\label{fig:ES7_3N8} (Color online) Entanglement spectrum for the Coulomb Hamiltonian as a function of $z$-component of angular  momentum $L_z^A$ for filling fraction $\nu = 1/3$ and particle number $N=8$ for $d/l$ = 0 (left panel), $d/l$ = 4 (middle panel), $d/l$ = 6 (right panel).  The suspected CFT levels consistent with the Laughlin model state for each $L_z^A$ are marked.}
\end{figure*} 

The entanglement gaps, which we denote as $\Delta_{i}$ for $i = \Delta L$, are calculated by finding the difference between the largest suspected low-lying CFT level and the next highest level at the given value of $L_z^A$ in the ES of the numerically obtained Coulomb ground states for varying $d/l$.  We also calculate the minimal gap between CFT and non-CFT levels for $\Delta L\leq m$, which we denote as $\Delta_{0-m}$ where $m$ is the cutoff described above.  The minimal gap gives us a qualitative measure of how well separated, overall, the CFT levels are from the generic non-CFT levels. The suspected low-lying CFT levels are identified by the expected counting described above.   An example of this procedure is shown in Fig.~\ref{fig:ESexample}, which shows the entanglement gaps in the ES of the $\nu = 1/3$ ground state for $d=0$ and $N=7$.  Note that throughout this work, when presenting figures showing ES, we color code the suspected CFT levels with red (gray) diamonds connected by a red (gray) dash and all other ES levels with a black dash.  The suspected CFT levels are chosen by calculating the ES for the model state (be it the Laughlin or the MR Pfaffian) and noting how many ES levels $n(L_z^A)$ there are for each $L_z^A$.  Then, when we consider the ES for the Coulomb Hamiltonians, we identify suspected CFT levels (and color code them) as the lowest $n(L_z^A)$ ES levels for 
each $L_z^A$.

Entanglement gaps as a function of finite layer thickness $d/l$ for the Laughlin filling fraction $\nu = 1/3$ are shown in Fig.~\ref{fig:ESGaps1_3}.  The entanglement gaps are slightly decreasing with $d/l$ for all cases, indicating that the states are weakening.  These trends are similar to those observed in the EE at $\nu = 1/3$.  Note that the minimal gap for $N=6$ and $N=8$ is initially small and becomes zero for $d/l\gtrsim4$.  This may indicate that the FQH state collapses at a finite thickness, as has been shown in previous works\cite{Zhang86,SongHe90} (the previous works showed the FQHE to collapse at very large $d/l$).  However, this effect is not seen in the minimal entanglement gap for $N=7$.  This ``even-odd'' finite size effect is likely due in part to a trade-off between the finite size cutoff and the number of available orbitals.  Indeed, the finite size cutoff is the same for $N=6$ and $N=7$, but the larger Hilbert space for the $N=7$ case allows for more ``edge excitations'' that strengthen each entanglement gap, not just the minimal gap, compared to $N=6$. 
In all cases, however, the overall trends in the entanglement gaps (i.e., slight decrease with $d/l$) are qualitatively similar to those seen in the EE (in particular, $\Delta S_E$) and the overlap in Refs. \onlinecite{PetersonPRL08,Peterson08}.

To illustrate this overall trend in the entanglement gaps for $\nu = 1/3$, we provide the ES of the ground states in Fig.~\ref{fig:ES1_3N8} for $d/l$ = 0, 2 and 6.  We have marked the levels that are consistent with the counting found in the ES of the Laughlin model state shown in Fig.~\ref{fig:ESLaughModel} for all values of $\Delta L$ and indicate our chosen finite size cutoff.  We see that qualitatively, the ES is largely insensitive to finite $d/l$.  Moreover, on the right of the finite size cutoff, except for the largest CFT state at $\Delta L = 4$, the low-lying CFT levels are well-separated from the higher energy generic levels.

We now examine the case when $\nu = 7/3$ in comparison.  In Fig.~\ref{fig:ESGaps7_3} are the entanglement gaps as a function of finite layer thickness.  For $N=6$, the root entanglement gap, $\Delta_0$, generally increases with $d/l$. $\Delta_1$ and $\Delta_2$ each have a weak, local maxima near $d/l \sim 4$ and $\Delta_3$ is actually decreasing with $d/l$.  Moreover the minimal entanglement gap is zero throughout. The entanglement gaps for $N=7$ are are each monotonically increasing with $d/l$, similarly to $\Delta_0$ in the $N=6$ case. The minimal gap, which is initially zero, opens at $d/l\sim3$ and then gradually increases with $d/l$ in this case. The case when $N=8$ shows trends similar to the $N=6$ case.  Here $\Delta_0$, $\Delta_1$, and $\Delta_2$ increase with $d/l$, $\Delta_3$ has a local maxima near $d/l\sim4$, and $\Delta_4$ decreases with $d/l$.  The minimal gap for $N=8$ is zero throughout. Again we see an ``even-odd'' finite size effect in the entanglement gaps as was seen with $\nu = 1/3$.  However, in this case, we have entanglement gaps that increase, decrease, or have a weak maxima as a function of $d/l$.  This is in contrast to the $\nu = 1/3$ case where \emph{all} entanglement gaps follow the same trend with finite $d/l$.  The different trends in the entanglement gaps may suggest that the topological signature of the $\nu = 7/3$ state differs from that of the Laughlin state. 

Some illustrative examples of ES at $\nu = 7/3$ are given in Fig.~\ref{fig:ES7_3N8} with $N = 8$ and $d/l$ = 0, 4 and 6.  The given ES appear to have structure similar to that seen in the $\nu = 1/3$ case, however, we see that for $\Delta L = 4$, the higher energy ``CFT'' states are virtually indistinguishable from the ``generic states''.  This ``blending'' appears to get worse for larger $d/l$.  Again, these results may suggest that the Laughlin model state is not an accurate description for the $\nu = 7/3$ state.

\begin{figure*}
\includegraphics[width=.65\textwidth]{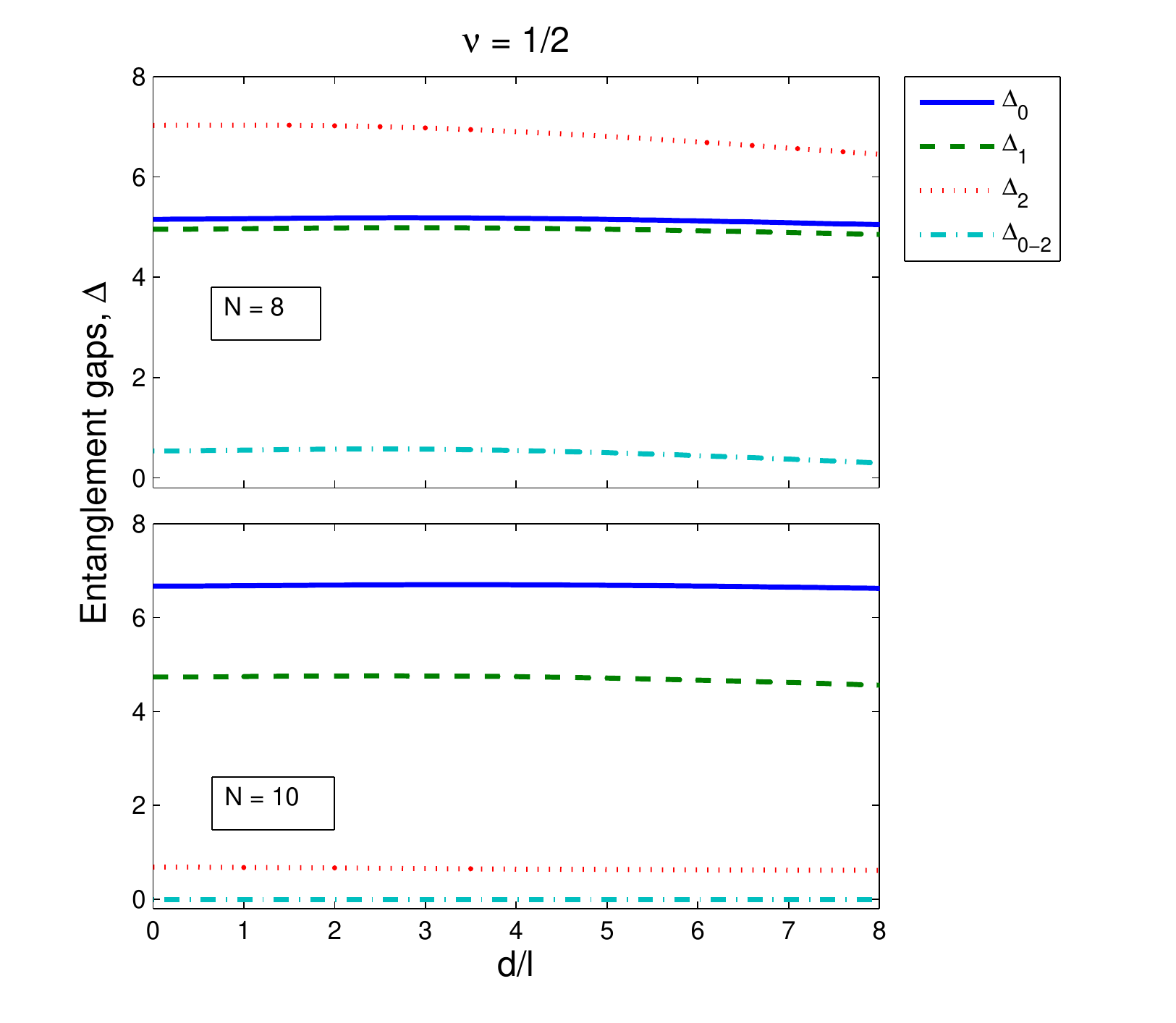}
\caption{\label{fig:ESGaps1_2} (Color online) Entanglement Gaps for the Coulomb Hamiltonian as a function of finite layer thickness, $d/l$ for filling fraction $\nu = 1/2$ and particle number $N = 8$ (top panel) and $N = 10$ (bottom panel) with partition at the equator. }
\end{figure*}

\begin{figure*}
\includegraphics[width=.9\textwidth]{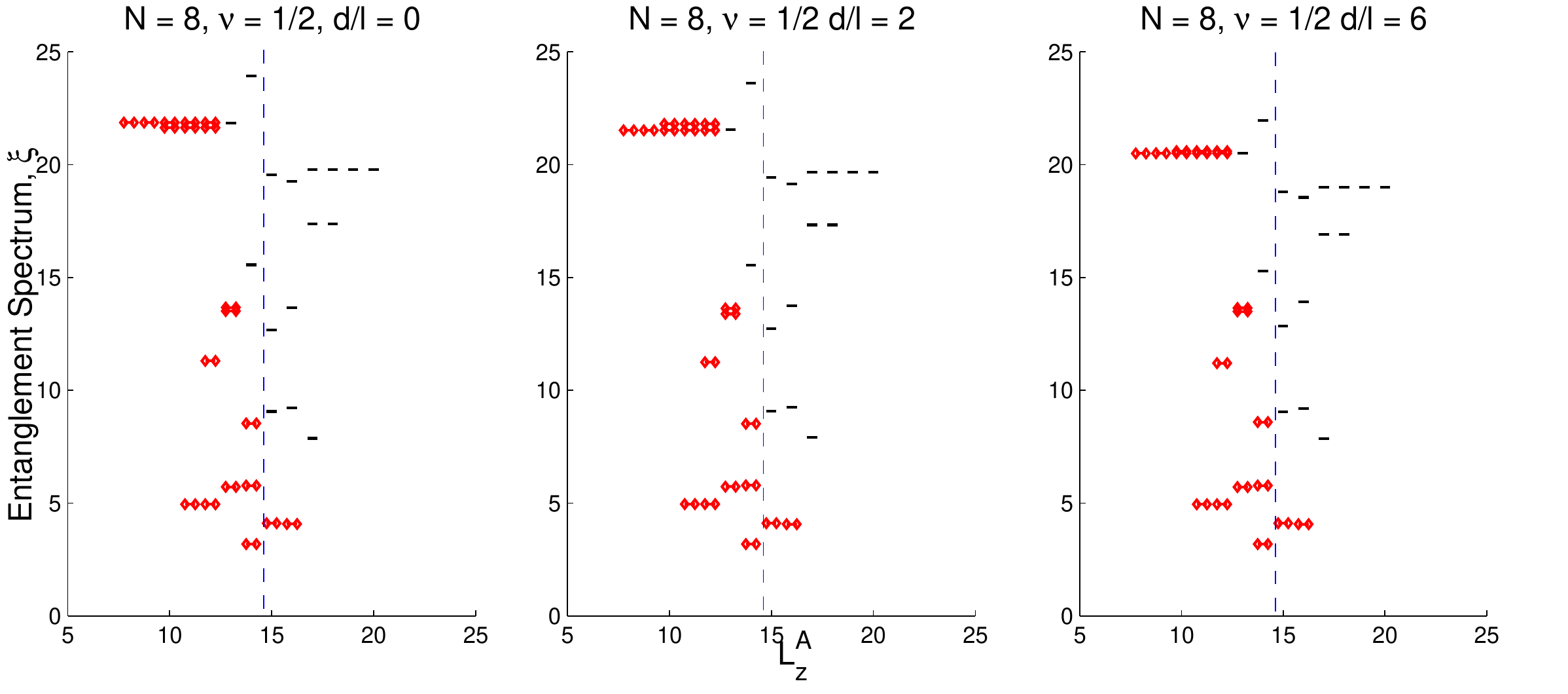}
\caption{\label{fig:ES1_2N8} (Color online) Entanglement spectrum for the Coulomb Hamiltonian as a function of $z$-component of angular  momentum $L_z^A$ for filling fraction $\nu = 1/2$ and particle number $N=8$ for $d/l$ = 0 (left panel), $d/l$ = 2 (middle panel), $d/l$ = 6 (right panel).  The suspected CFT levels consistent with the MR Pfaffian model state for each $L_z^A$ are marked.}
\end{figure*} 

\begin{figure*}
\includegraphics[width=.65\textwidth]{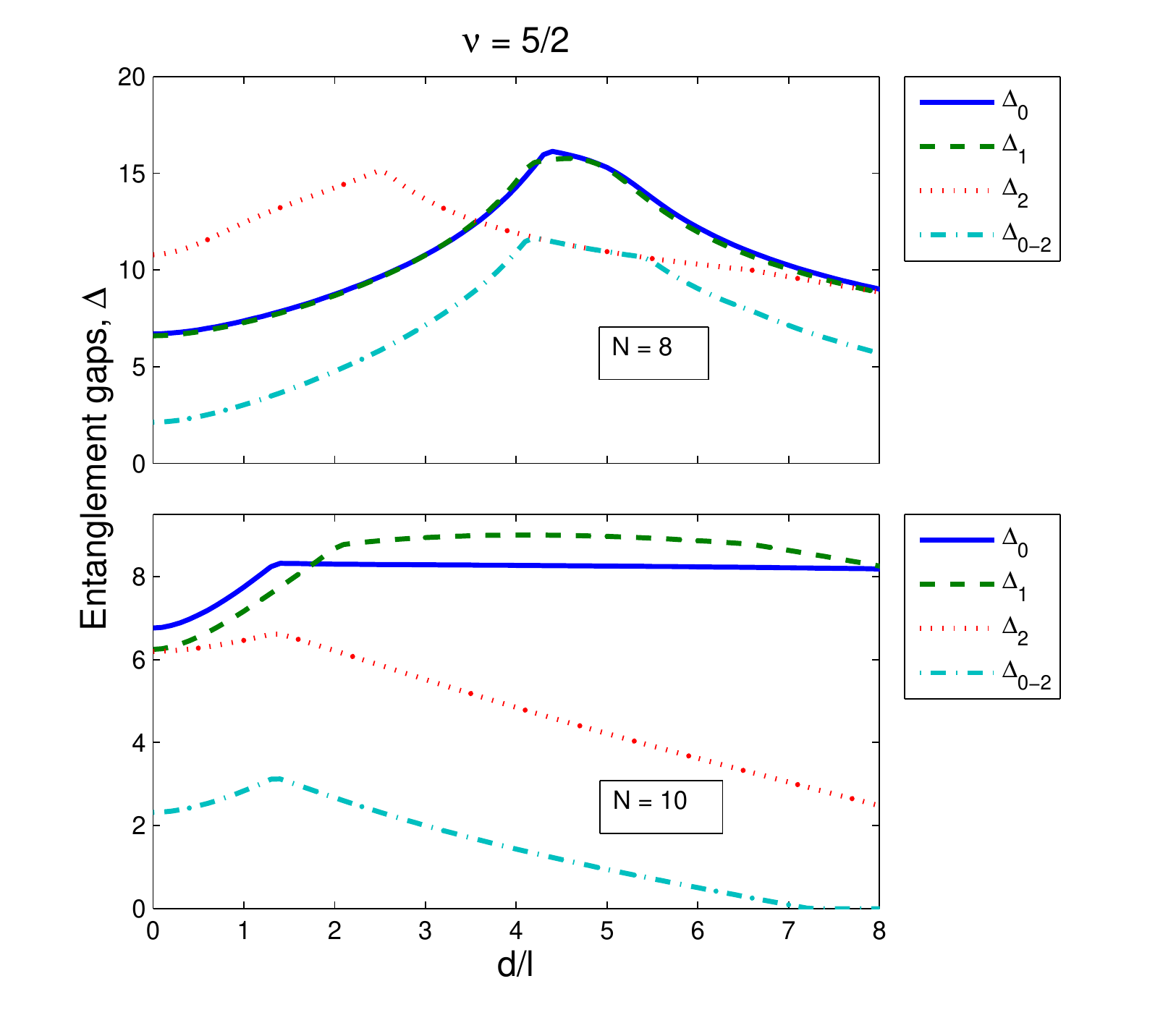}
\caption{\label{fig:ESGaps5_2} (Color online) Entanglement Gaps for the Coulomb Hamiltonian as a function of finite layer thickness, $d/l$ for filling fraction $\nu = 5/2$ and particle number $N = 8$ (top panel) and $N = 10$ (bottom panel) with partition at the equator. }
\end{figure*}

\begin{figure*}
\includegraphics[width=.9\textwidth]{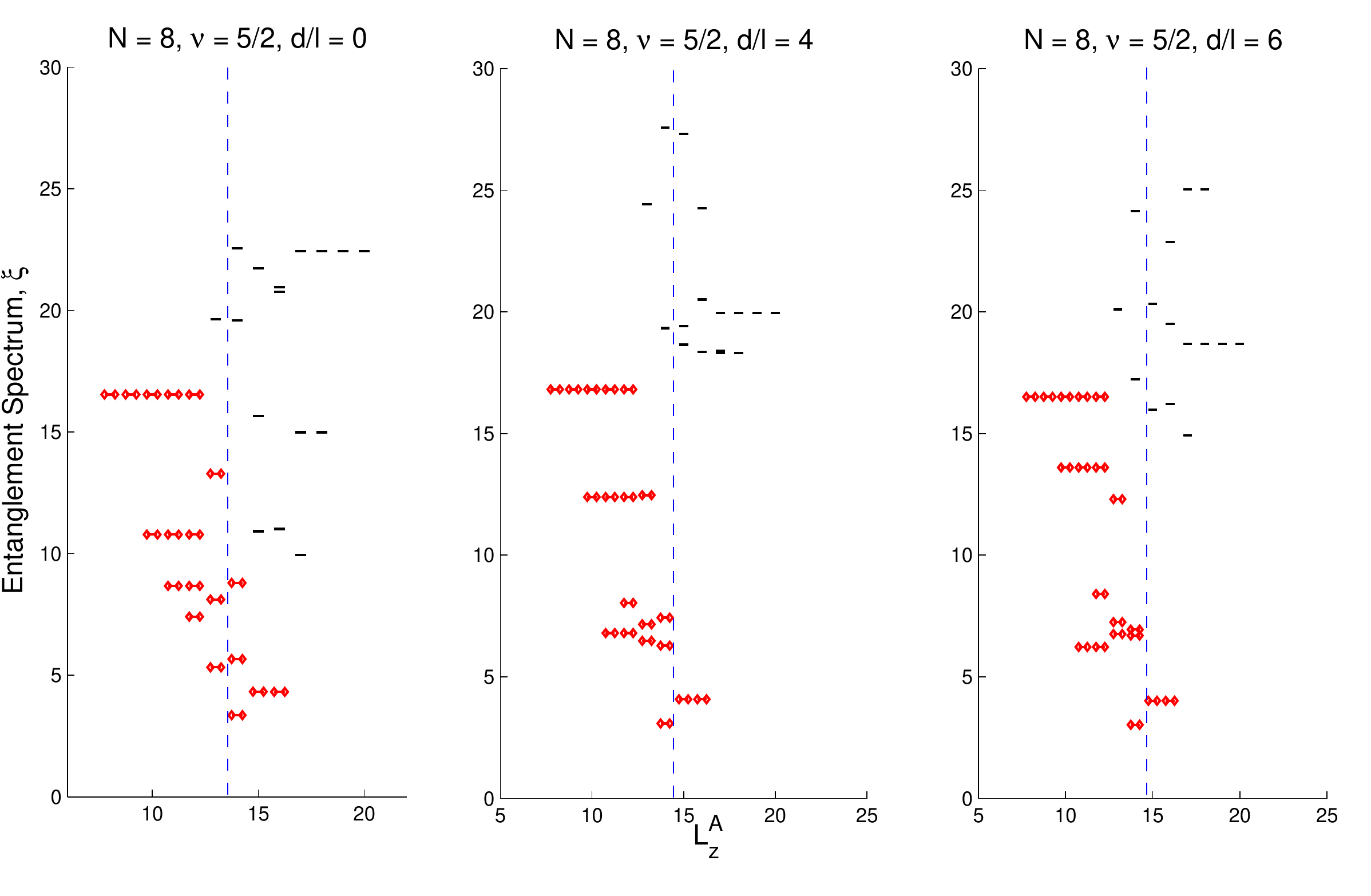}
\caption{\label{fig:ES5_2N8} (Color online) Entanglement spectrum for the Coulomb Hamiltonian  as a function of $z$-component of angular of angular momentum $L_z^A$ for filling fraction $\nu = 5/2$ and particle number $N=8$ for $d/l$ = 0 (left panel), $d/l$ = 4 (middle panel), $d/l$ = 6 (right panel).  The suspected CFT levels consistent with the MR Pfaffian model state for each $L_z^A$ are marked.}
\end{figure*}

Results on the entanglement gaps for the even denominator filling fraction $\nu = 1/2$ are shown in Fig.~\ref{fig:ESGaps1_2}.  Here we see that the entanglement gaps slightly decrease with $d/l$ and behave similarly to the EE at this filling fraction.  Also note that for $N=8$, the minimal gap is small and decreases with $d/l$, while for $N=10$, the minimal gap is zero throughout.  As mentioned earlier, there has been no definitive experimental observation of FQHE at $\nu=1/2$ in monolayer systems consistent 
with our calculations.  The full ES of the ground states with $\nu=1/2$ are given in Fig.~\ref{fig:ES1_2N8} for $d/l$ = 0, 2 and 6 and $N=8$. Qualitatively, we see that the ES is largely insensitive to the finite-thickness effect.  Moreover, the largest suspected CFT level for $\Delta L = 2$ is well separated from the other CFT levels and appears to be more consistent with the generic levels.  Again, this suggests that $\nu = 1/2$ is not described by the MR Pfaffian wavefunction.

Fig.~\ref{fig:ESGaps5_2} shows the entanglement gaps at filling fraction $\nu = 5/2$.  For $N=8$ each entanglement gap peaks at a certain value for $d/l$. In particular $\Delta_2$ peaks near $d/l\sim 2.5$; the other gaps peak near $d/l\sim4$.  We also see peaks in the entanglement gaps for the case when $N=10$.  Here, the gaps gradually rise to a local maxima near $d/l\sim 1.5$ and then slowly decay for increasing $d/l$.  Note that the gaps in this case are generally smaller compared to those observed for $N=8$.  These results may suggest that there is a slight difference in the finite-size effect on the different MR Pfaffian CFT sectors.  However, these results are qualitatively similar to the EE results and the results on the overlap in Refs. \onlinecite{Peterson08} (i.e., the MR Pfaffian signature of the $\nu = 5/2$ state is strengthened by the finite size effect). 

We also provide the ES of the $\nu = 5/2$ state for $N=8$ in Fig.~\ref{fig:ES5_2N8} for $d/l = 0$, 4, and 6.  Here, we see the ES ``opens'' at $d/l = 4$, giving a larger separation between the CFT and generic levels in the spectrum compared to $d/l$ = 0 and 6.  Again, these results suggests the $\nu = 5/2$ is, indeed, described by the MR Pfaffian wavefunction, and this description is more stable at finite thickness.

In summary, the entanglement gaps in the ES have a similar dependence on finite thickness as the EE, leading to similar conclusions.  However finite size effects prevent us from making definitive statements.  In the next section, we attempt to alleviate this problem using the conformal limit. 

\subsection{The Conformal Limit} \label{sec:CL}
\begin{figure*}
\includegraphics[width=.9\textwidth]{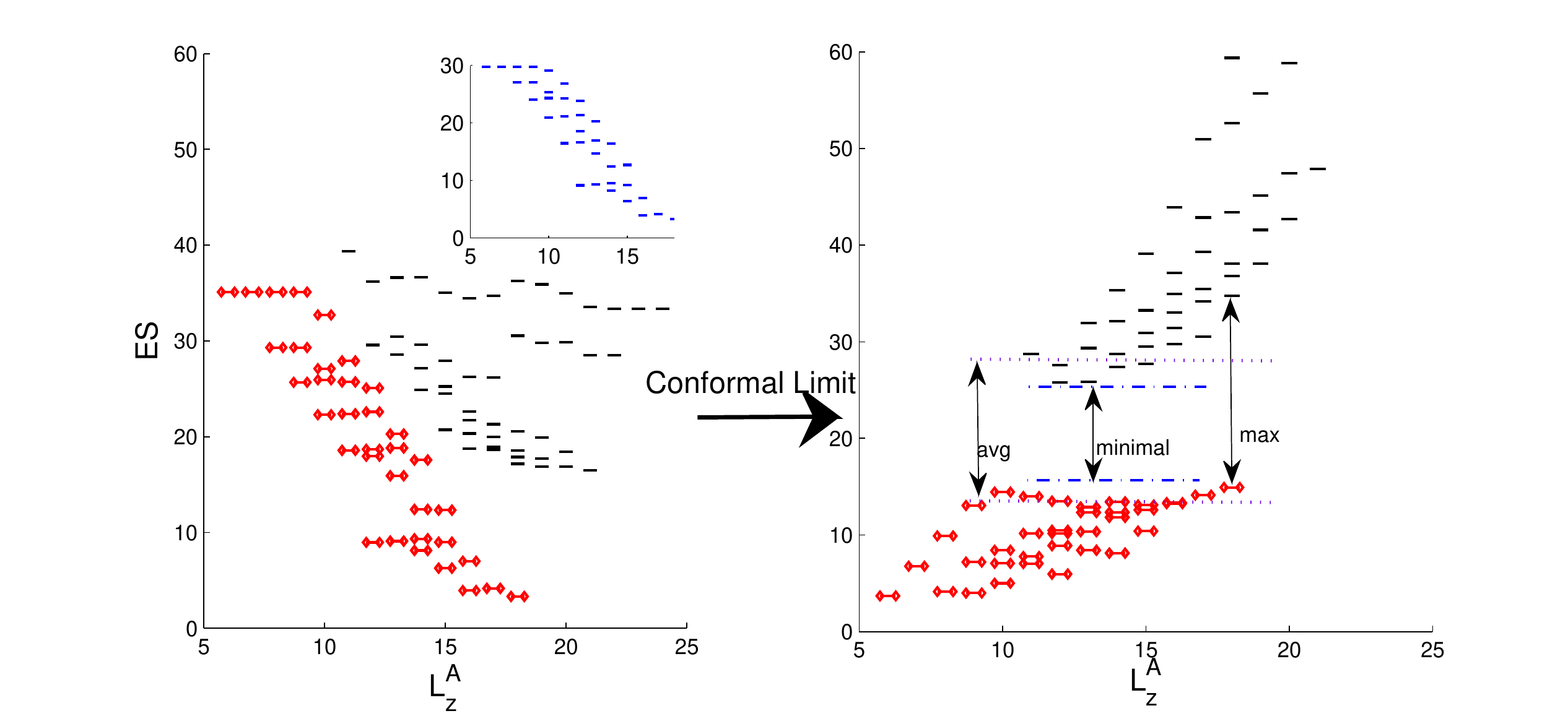}
\caption{\label{fig:CLdemo} (Color online) ES for the Coulomb Hamiltonian for $\nu = 1/3$, $d=0$, and $N=7$ as a function of $z$-component of angular momentum, $L_z^A$ before and after taking the CL.  The suspected CFT states are based on the ES of the Laughlin model wave function shown (inset) and are marked in the ES.  Illustration for the minimal, maximum, and average entanglement gap is also shown.}
\end{figure*}

\begin{figure*}
\includegraphics[width=.6\textwidth]{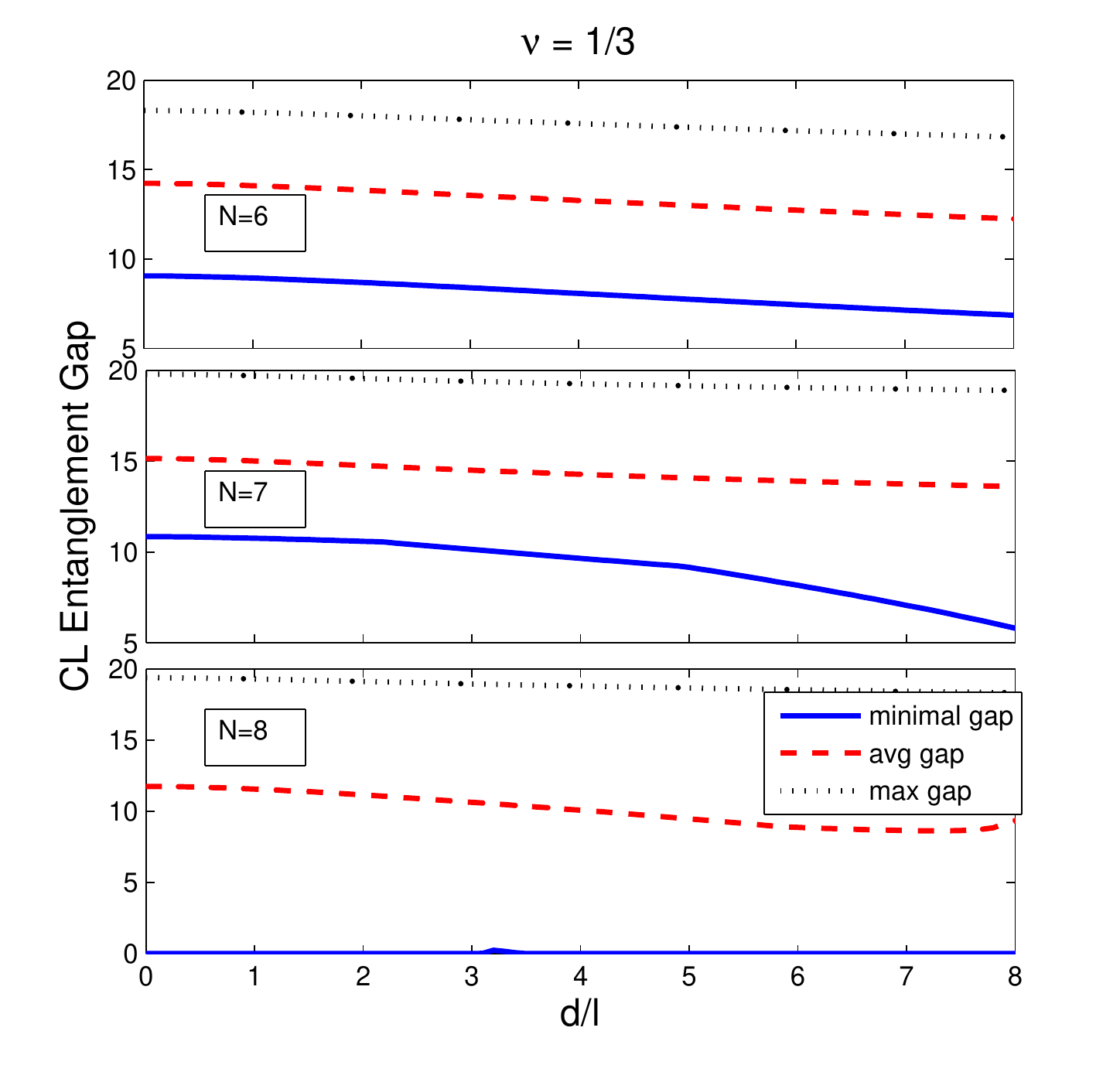}
\caption{\label{fig:CLES1_3} (Color online) Conformal limit entanglement gaps for the Coulomb Hamiltonian as a function of finite layer thickness, $d/l$ for filling fraction $\nu = 1/3$ and particle number $N = 6$ (top panel), $N = 7$ (middle panel), and $N = 8$ (bottom panel). }
\end{figure*}

\begin{figure*}
\includegraphics[width=.9\textwidth]{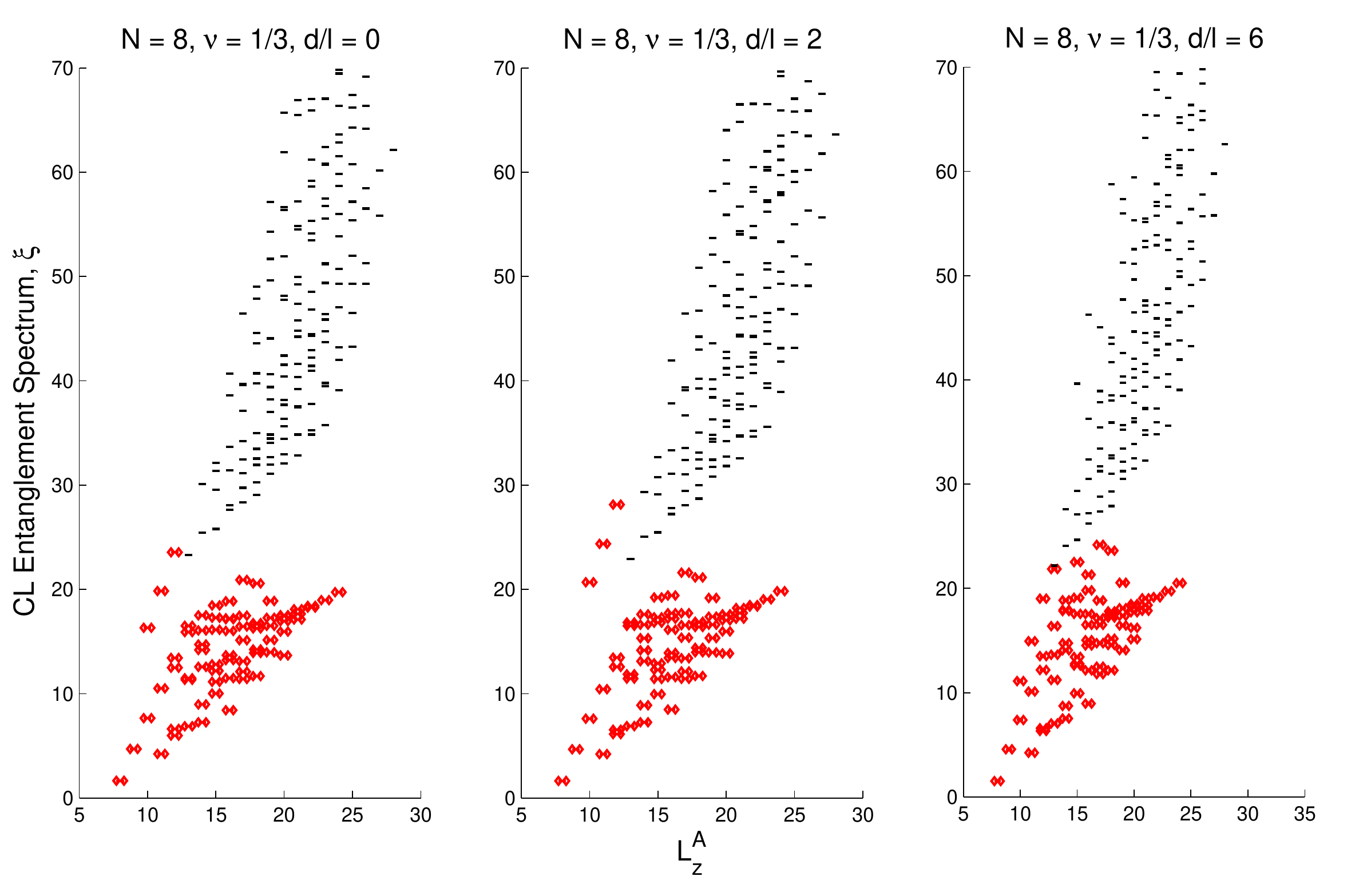}
\caption{\label{fig:CLES1_3N8} (Color online) Conformal limit entanglement spectrum for the Coulomb Hamiltonian as a function of $z$-component of angular momentum $L_z^A$ for filling fraction $\nu = 1/3$ and particle number $N=8$ for $d/l$ = 0 (left panel), $d/l$ = 2 (middle panel), $d/l$ = 6 (right panel).  The suspected CFT levels consistent with the Laughlin model state for each $L_z^A$ are marked.}
\end{figure*}

\begin{figure*}
\includegraphics[width=.6\textwidth]{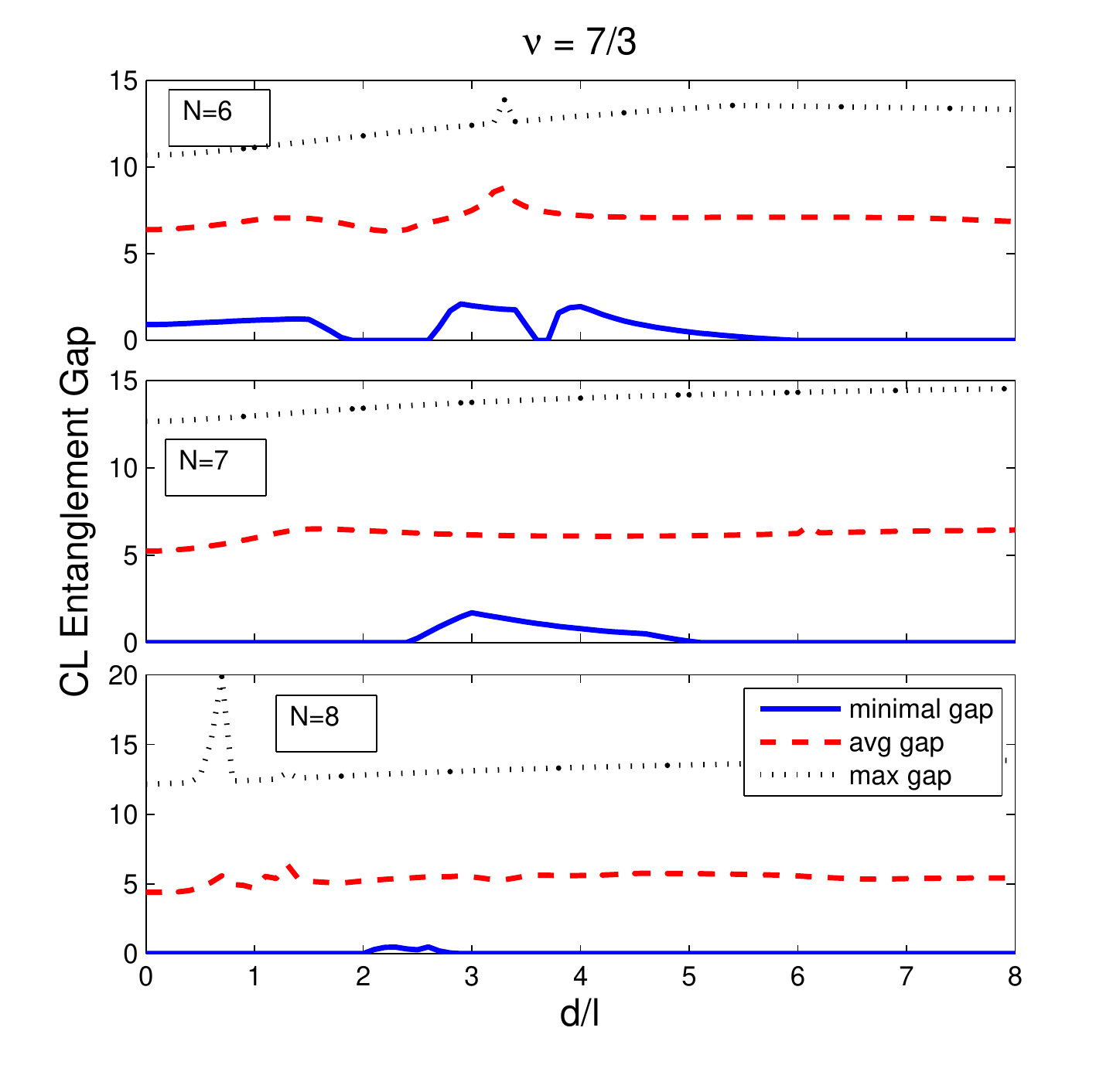}
\caption{\label{fig:CLES7_3} (Color online) Conformal limit entanglement gaps for the Coulomb Hamiltonian  as a function of finite layer thickness, $d/l$ for filling fraction $\nu = 1/3$ and particle number $N = 6$ (top panel), $N = 7$ (middle panel), and $N = 8$ (bottom panel). }
\end{figure*}

\begin{figure*}
\includegraphics[width=.9\textwidth]{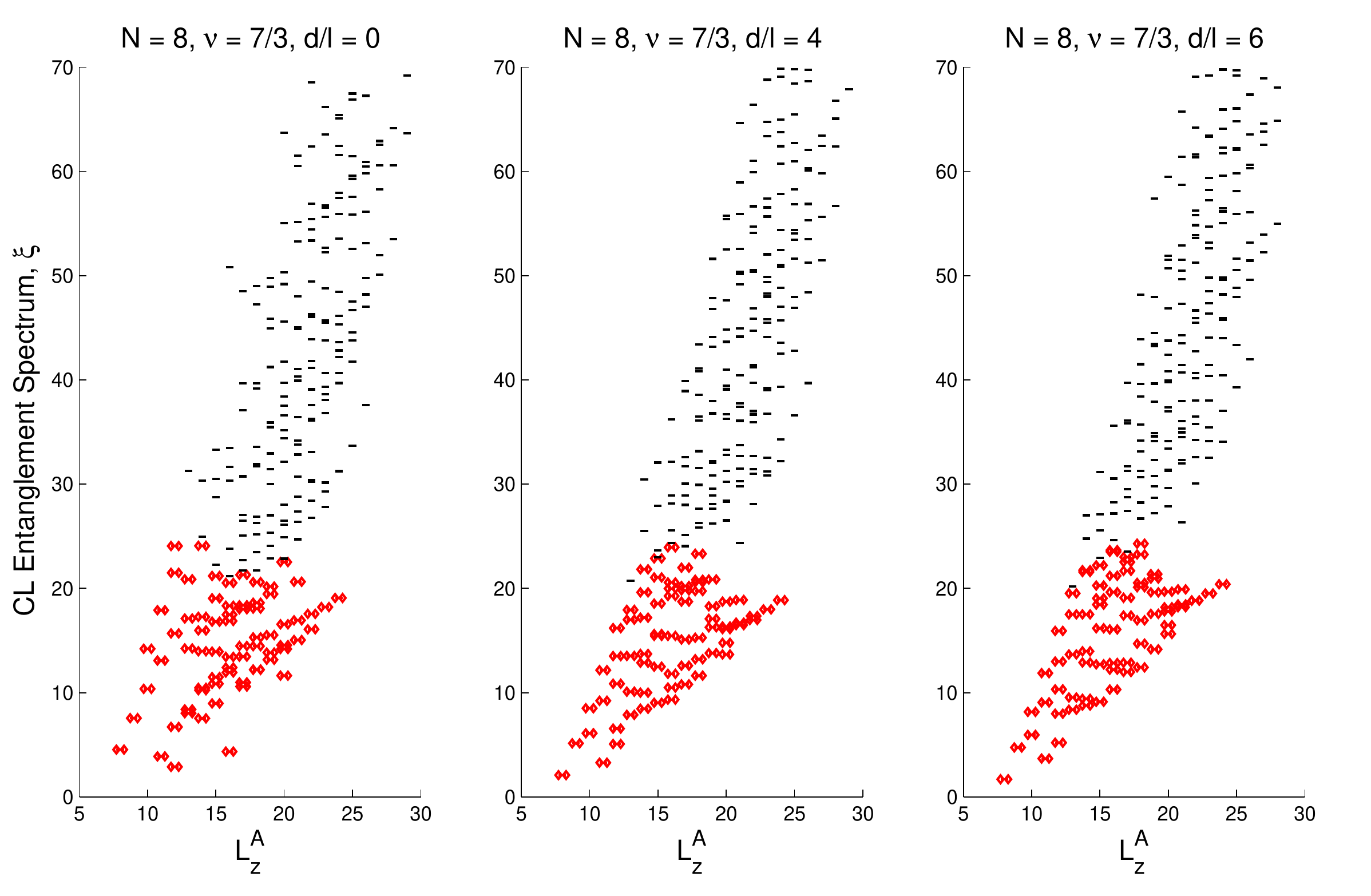}
\caption{\label{fig:CLES7_3N8} (Color online) Conformal Limit Entanglement spectrum for the Coulomb Hamiltonian as a function of $z$-component of angular momentum $L_z^A$ for filling fraction $\nu = 1/3$ and particle number $N=8$ for $d/l$ = 0 (left panel), $d/l$ = 4 (middle panel), $d/l$ = 6 (right panel).  The suspected CFT levels consistent with the Laughlin model state for each $L_z^A$ are marked.}
\end{figure*}

In the previous section, we used entanglement gaps in the ES to evaluate the ``strength" of a state as a function of $d/l$ and we were able to confirm the MR Pfaffian signature of the $\nu = 5/2$ state and distinguish it from the (lack of) signature of the $\nu = 1/2$ state.  However, we have intentionally ignored a significant part of the ES in order to avoid finite size effects, i.e., we focused on the 
region of the ES with small $\Delta L$ (see Fig.~\ref{fig:ESexample}).  We determined the size of this region by examining where the ES of the finite sized MR Pfaffian and Laughlin model states deviate from the conjectured structure in the thermodynamic limit (i.e., the edge state level counting given by the suspected CFT).  Thus, we have confirmed the MR Pfaffian and Laughlin signatures only to a certain extent because, in fact, there is not an actual entanglement gap in the ES.  

It has been conjectured that the \emph{full} entanglement spectrum of the finite sized model states contain information on the topological signature of the FQH state.\cite{Hermanns10, Chandran11}  Thus, all states in the ES can be used to identify the topological quantities.  With this in mind, we now examine the entanglement spectrum of quasi-2D FQH states in the ``conformal limit" (CL), which reportedly allows the use of the entire spectrum to examine the state by unambiguously defining a full entanglement gap.  As discussed briefly above, and at length by Thomale \textit{et al.} in Ref.~\onlinecite{Thomale10}, the CL works by removing finite size effects due to the curvature of the sphere and gives an ES with a ``full" unambiguous entanglement gap in the spectrum for 
topologically ordered states.  Thus the presence of an entanglement gap in the conformal limit is conjectured to be a sign of topological order.  A demonstration of an ES before and after the CL is given in Fig.~\ref{fig:CLdemo}.  After taking the CL of an ES (CLES), we determine the ``minimal gap" by taking the difference between the highest suspected CFT level and the lowest generic level in the entire spectrum.  The suspected CFT levels are determined by comparing the CLES to that of the model state with the assumption that \emph{all} levels in the ES of the model state are CFT levels.  For comparison, we examine the entanglement gaps for each value of $L_z^A$ and define the ``average gap" as the average of the individual entanglement gaps. We also define the ``maximum gap" as the maximum of the entanglement gaps.  Individual gaps that are near infinite (i.e., no levels above the highest CFT level) are ignored.  The minimal gap, the average gap, and the maximum gap are calculated for each CLES as a function of the finite layer thickness, $d/l$.

CLES entanglement gaps as a function of finite layer thickness $d/l$ for $\nu=1/3$ (LLL) are shown in Fig.~\ref{fig:CLES1_3}.  For $N=6$ and 7, the entanglement gap measures decreases with $d/l$ but remains finite throughout.  This behavior is qualitatively similar to the ES gaps for small $\Delta L$, as well as the EE results, suggesting a weakening of the Laughlin state.  The fact that the minimal entanglement gap in the ES for the $N=6$ case (Fig.~\ref{fig:ESGaps1_3}) differs from the minimal gap in the CLES may indicate that the closing of the gap in the ES is due to finite size effects related to the curvature of the geometry rather than the limited number of LL orbitals.  However, the minimal gap for the case where $N=8$ seems anomalous.  Although the average and maximum gaps follow similar qualitative trends, the minimal gap is at or near 0 for all values of $d/l$, including $d=0$.  How to interpret this result is unclear since there is a general consensus that the Laughlin state does, indeed, model the $\nu=1/3$ state.  We can shed some light on this anomaly by examining the CLES of the FQH states directly.  Fig.~\ref{fig:CLES1_3N8} shows the CLES for the $\nu = 1/3$, $N=8$ FQH state at finite thicknesses $d/l = 0$, 2 and 6.  The suspected CFT levels are marked in each plot.  

We examine CLES in the SLL case ($\nu = 7/3$) in Fig.~\ref{fig:CLES7_3}.  In general, the behavior of each gap measure differs with varying $d/l$.  The minimal gap appears fragile and virtually disappears for larger $N$.  The average gap has two local maxima in $d/l$ for $N=6$.  Only one of the local maxima in the average gap is preserved when we look at the $N=7$ case, and for $N=8$, the average gap fluctuates.   The maximum gap, in general, increases with increasing $d/l$ but has a notable peak near $d/l \sim 0.7$ for $N=8$.  The inconsistency in these results may suggest, from the ES and EE results, that the Laughlin model state is not a suitable model for the $\nu = 7/3$ state, or other ignored effects are needed for the Coulomb state to be adequately described by the Laughlin 
state.

Fig.~\ref{fig:CLES7_3N8} shows the CLES results for $\nu = 7/3$, $N=8$ at finite thickness $d/l$ = 0, 4, and 6.  We note that for each value of $d/l$, there is very little separation between the suspected CFT levels and the generic levels.  Indeed, if the suspected CFT levels were not marked, there is no clear entanglement gap across the whole spectrum.  However, there does appear to be structure in the CLES for small values of $\Delta L$ (i.e., near the ``root'' configuration).  What this may imply about the topological signature of the $\nu = 7/3$ state is not clear.    

\begin{figure*}
\includegraphics[width=.6\textwidth]{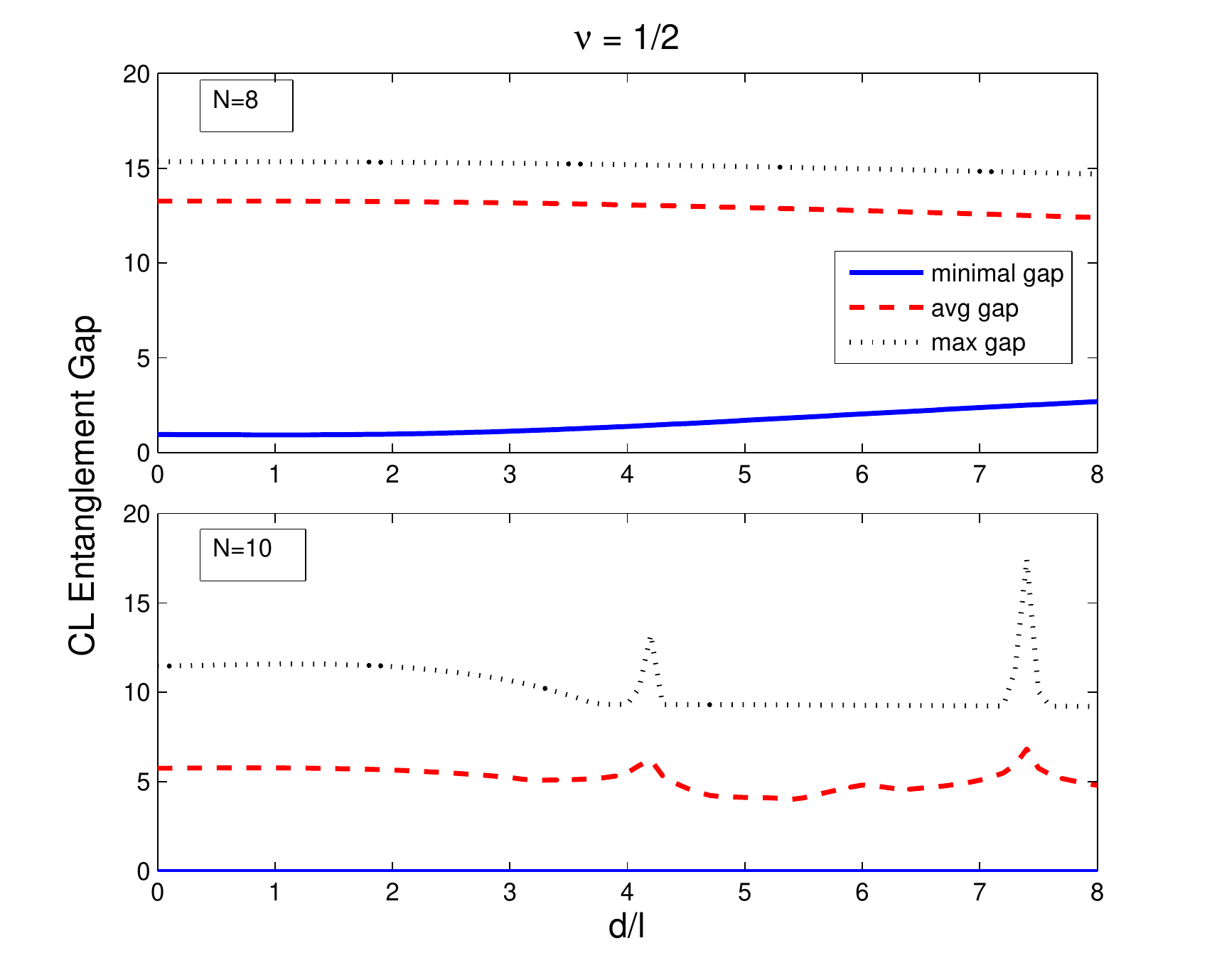}
\caption{\label{fig:CLES1_2} (Color online) Conformal Limit Entanglement Gaps as a function of finite layer thickness, $d/l$ for filling fraction $\nu = 1/2$ and particle number $N = 8$ (top panel) and $N = 10$ (bottom panel) }
\end{figure*}

\begin{figure*}
\includegraphics[width=.9\textwidth]{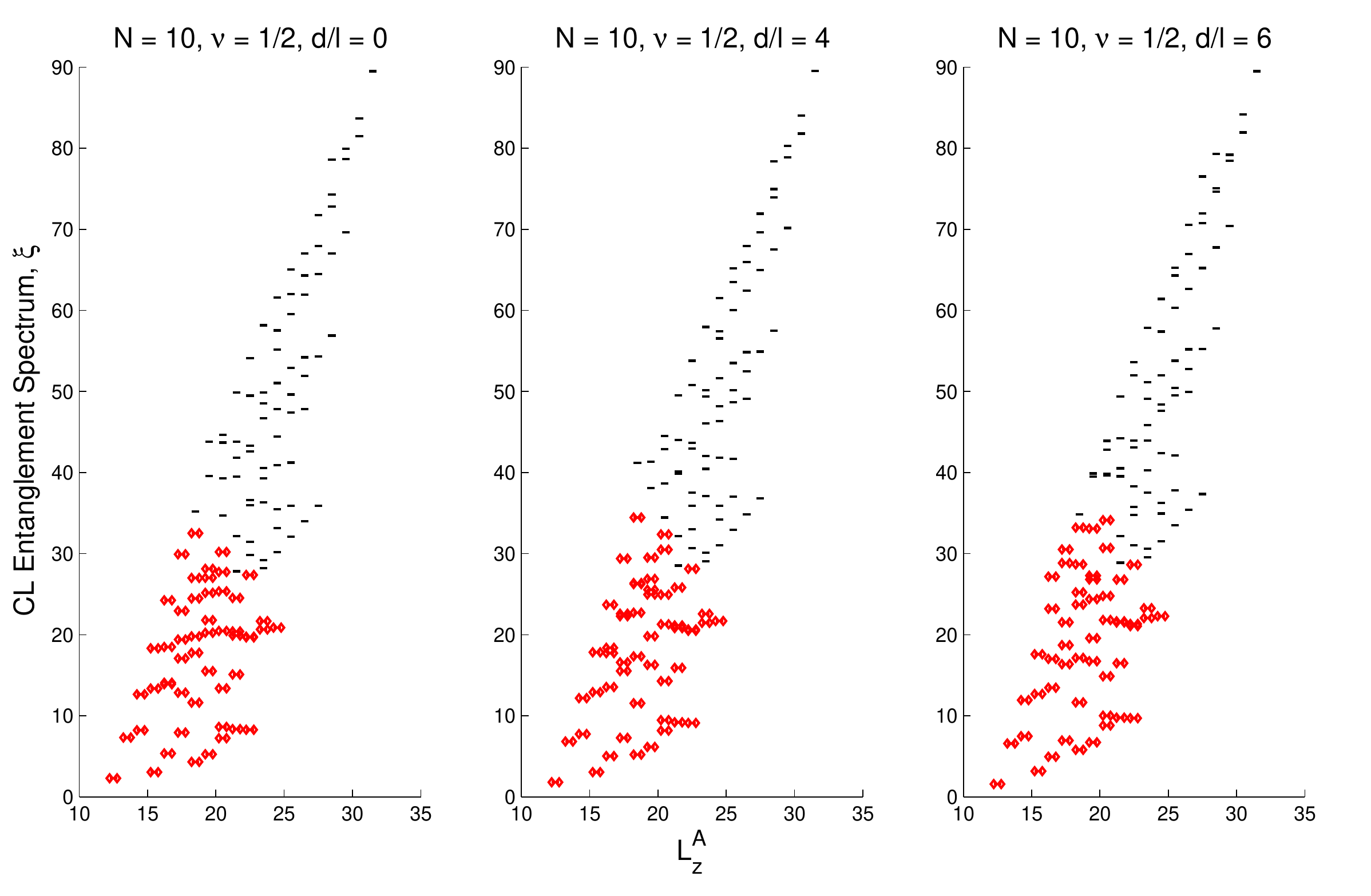}
\caption{\label{fig:CLES1_2N10} (Color online) Entanglement spectrum as a function of $z$-component of angular momentum $L_z^A$ for filling fraction $\nu = 1/2$ and particle number $N=10$ for $d/l$ = 0 (left panel), $d/l$ = 4 (middle panel), $d/l$ = 6 (right panel).  The suspected CFT levels consistent with the MR Pfaffian model state for each $L_z^A$ are marked.}
\end{figure*} 

\begin{figure*}
\includegraphics[width=.6\textwidth]{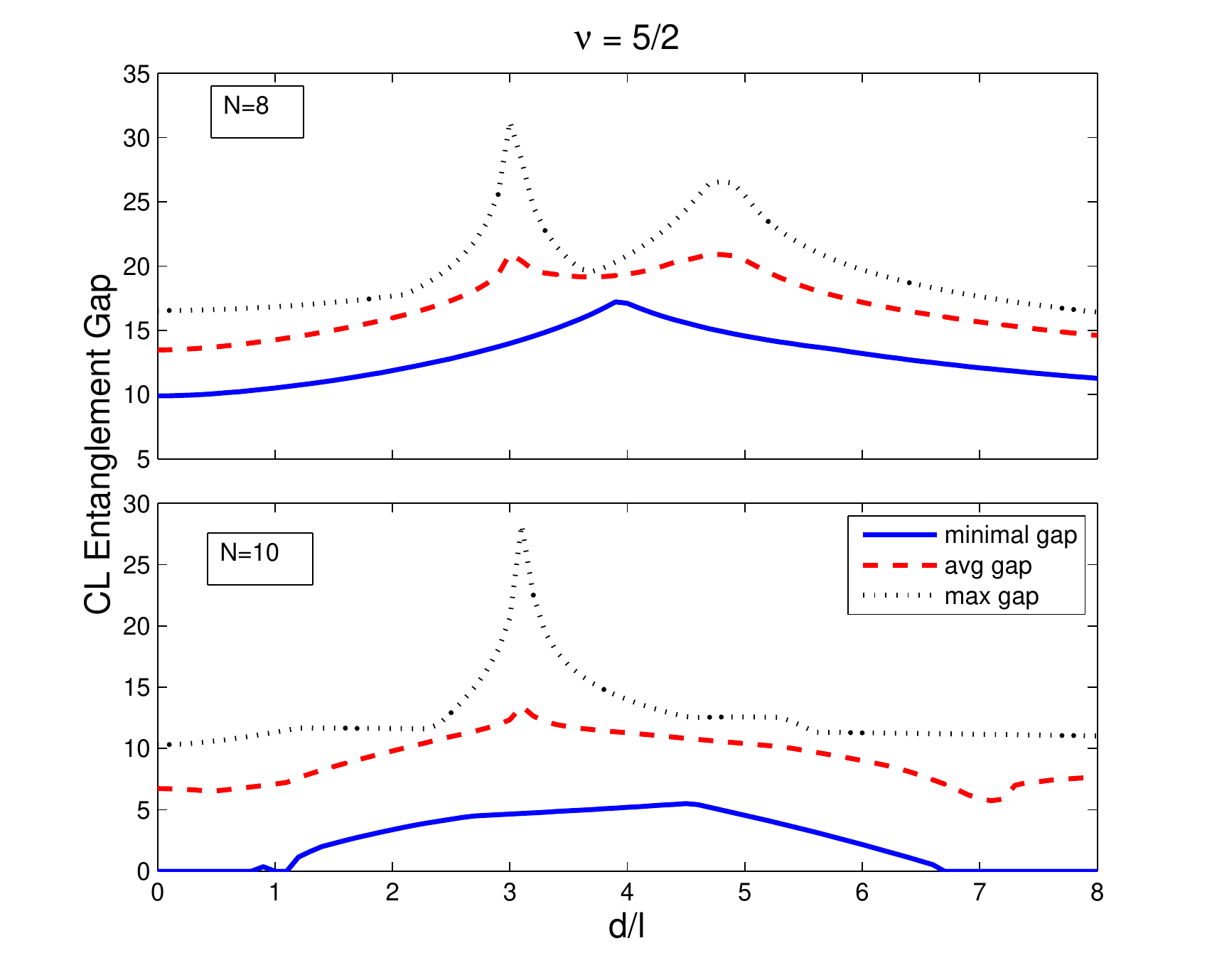}
\caption{\label{fig:CLES5_2} (Color online) Entanglement Gaps as a function of finite layer thickness, $d/l$ for filling fraction $\nu = 5/2$ and particle number $N = 8$ (top panel) and $N = 10$ (bottom panel) with partition at the equator. }
\end{figure*}

\begin{figure*}
\includegraphics[width=.9\textwidth]{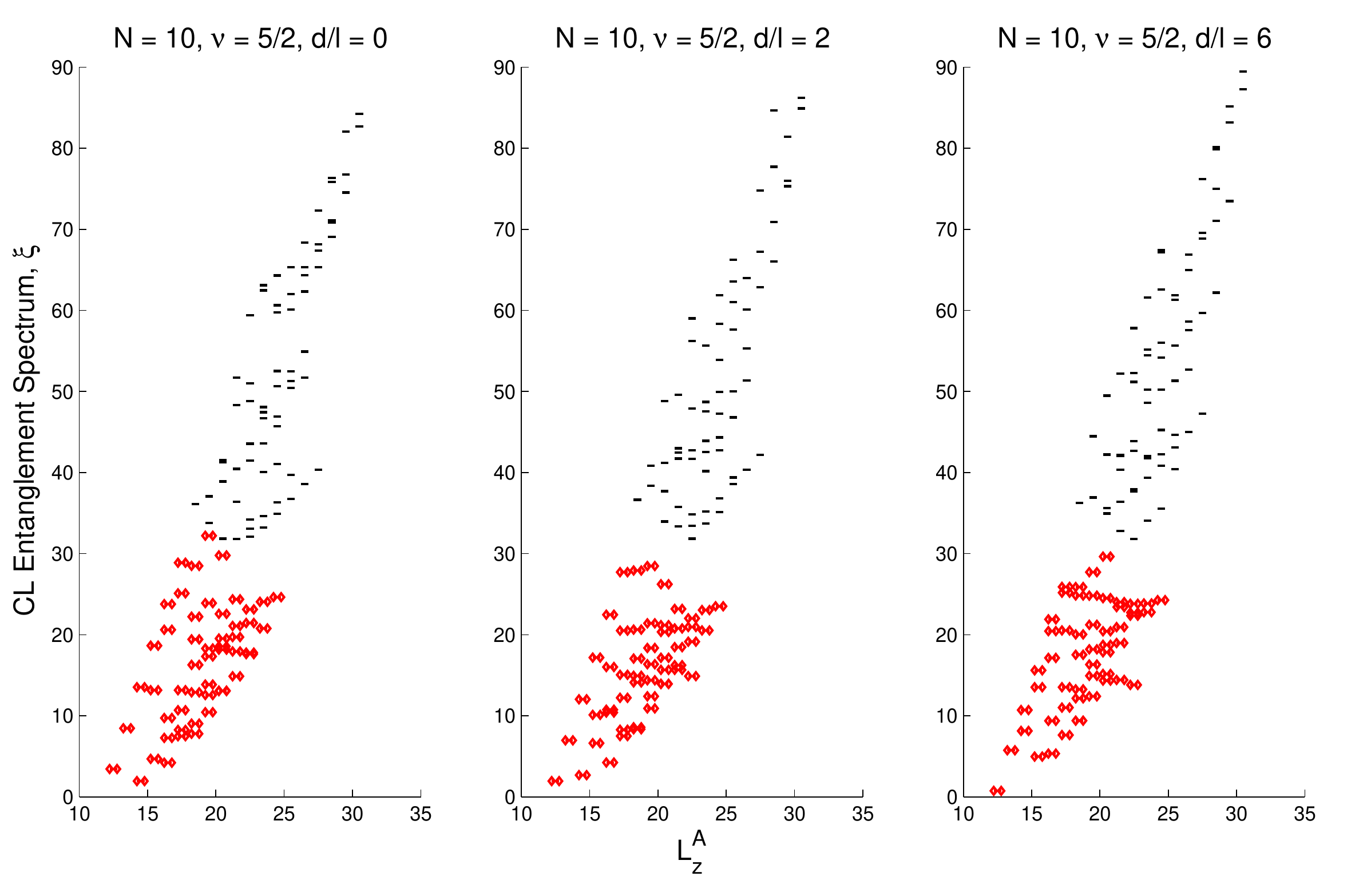}
\caption{\label{fig:CLES5_2N10} (Color online) Entanglement spectrum as a function of $z$-component of angular momentum $L_z^A$ for filling fraction $\nu = 5/2$ and particle number $N=10$ for $d/l$ = 0 (left panel), $d/l$ = 2 (middle panel), $d/l$ = 6 (right panel).  The suspected CFT levels consistent with the MR Pfaffian model state for each $L_z^A$ are marked.}
\end{figure*}

Results for the CLES gap measures in the even denominator $\nu = 1/2$ FQH state are shown in Fig.~\ref{fig:CLES1_2}.  We see that the minimal gap is nonzero and gradually increases with $d/l$ for $N=8$.  However, for $N=10$ the minimal gap is zero throughout.  The maximum and average gaps decrease with $d/l$ for $N=8$.  For $N=10$, the average gap has several local maxima, while the maximum gap decreases then suddenly becomes constant with $d/l$ with two sharp peaks.  Given our results on the EE and the ES for this state, the inconsistency between the $N=8$ and $N=10$ in the CLES gap measures may suggest that the MR Pfaffian model state is not a suitable model for $\nu=1/2$.  We also provide the CLES of the ground states in Fig.~\ref{fig:CLES1_2N10} for $N=10$ and $d/l = 0$, 4, and 6.  Qualitatively, the CLES do not change very much as a function of $d/l$, and there is no clear separation between the CFT and generic levels.  This, again, suggests that there is no FQH state at this filling fraction.

The CLES gap measures for $\nu = 5/2$ in the SLL are given in Fig.~\ref{fig:CLES5_2}.  For $N=8$, the minimal gap has a very pronounced peak near $d/l\approx4$.  The average and maximum gaps, however, have local minima near where the minimal gap is maximum.  These ``cusps'' are a result of level crossings.   For $N=10$, the minimal gap is initially zero, but becomes finite for non-zero $d/l$ and peaks near $d/l\approx4.5$.   The average and maximum gap in this case have similar shapes with a peak near $d/l\approx3$.  These results are qualitatively similar to the results of the ES, EE, and the overlap in Refs.~\onlinecite{PetersonPRL08,Peterson08}.  Moreover, the difference between $N=8$ and $N=10$ may suggest that finite thickness affects each sector of the suspected CFT differently, but larger system sizes are necessary to verify this.
In the CLES plots shown in Fig.~\ref{fig:CLES5_2N10} for $N=10$ and $d/l = 0$, 4 and 6 respectively, we see the entanglement gap between CFT and generic levels ``open'' at finite $d/l = 4$ compared to $d/l = 0$ and 6.  These results are consistent with results observed with the EE and the ES, indicating that MR Pfaffian signature strengthens with a finite $d/l$.

In summary, taking the conformal limit of the entanglement spectra provides us with a full entanglement gap in most cases with a finite thickness dependence that is qualitatively similar to the results on the EE.  The notable exceptions are the $\nu = 1/2$ which has little or no entanglement gap consistent with experimental observations, the $\nu = 1/3$ case at $N = 8$ that is not expected given results with other system sizes, and the $\nu = 7/3$ case that is consistent with the results on the EE and ES suggesting that other physics besides the Laughlin state alone is needed to explain this FQHE.  The case with $N = 8$ and $\nu = 1/3$, however, is inconsistent with most theory and experiment, but when we examine the spectra directly, there are a few ``spurious" states that cross an otherwise full gap.   The origin of these ``spurious" states are related to our use of planar Haldane pseudopotentials rather than spherical pseudopotentials and is discussed in the appendix.  However we do not expect this choice to alter the topological features of the state.  Therefore, this result may suggest that a ``full" quantitative entanglement gap is not necessary to identify a topological state.  In the next section, we introduce the concept of a entanglement spectral density of states where a qualitative, ``soft" gap may be identified in such cases.

\subsection{Entanglement Spectrum Density of States} \label{sec:DOS}

In the entanglement results presented above, we require a model state wavefunction for comparison in order to systematically define and calculate the entanglement gaps.  These methods have the obvious disadvantage of requiring an ansatz for comparison.  In the conformal limit case, we assume the low-lying states in the entanglement spectrum should have the exact counting as seen in the model entanglement spectrum.  This assumption may be premature since other finite size effects may cause the counting to deviate, even after taking the conformal limit especially at the largest $\Delta L$, see Fig.~\ref{fig:CLES1_3N8}.  With this in mind, we attempt to obtain a general qualitative sense for how the entanglement spectra vary with finite layer thickness by extending the analogy with ``energy levels" a bit further by calculating the ``density of entanglement spectral states."   With the density of states, we can qualitatively look for entanglement gaps without relying on a model state for comparison.  Also, we may be able to detect ``soft" gaps where a small number of states may be present within an otherwise prominent gap between two peaks in the density of states.   Thus in this section we briefly examine this extension by providing results for the density of states (DOS) of the entanglement spectrum, both with and without the conformal limit, as a function of finite layer thickness, $d/l$.

\begin{figure}
\includegraphics[width=.47\textwidth]{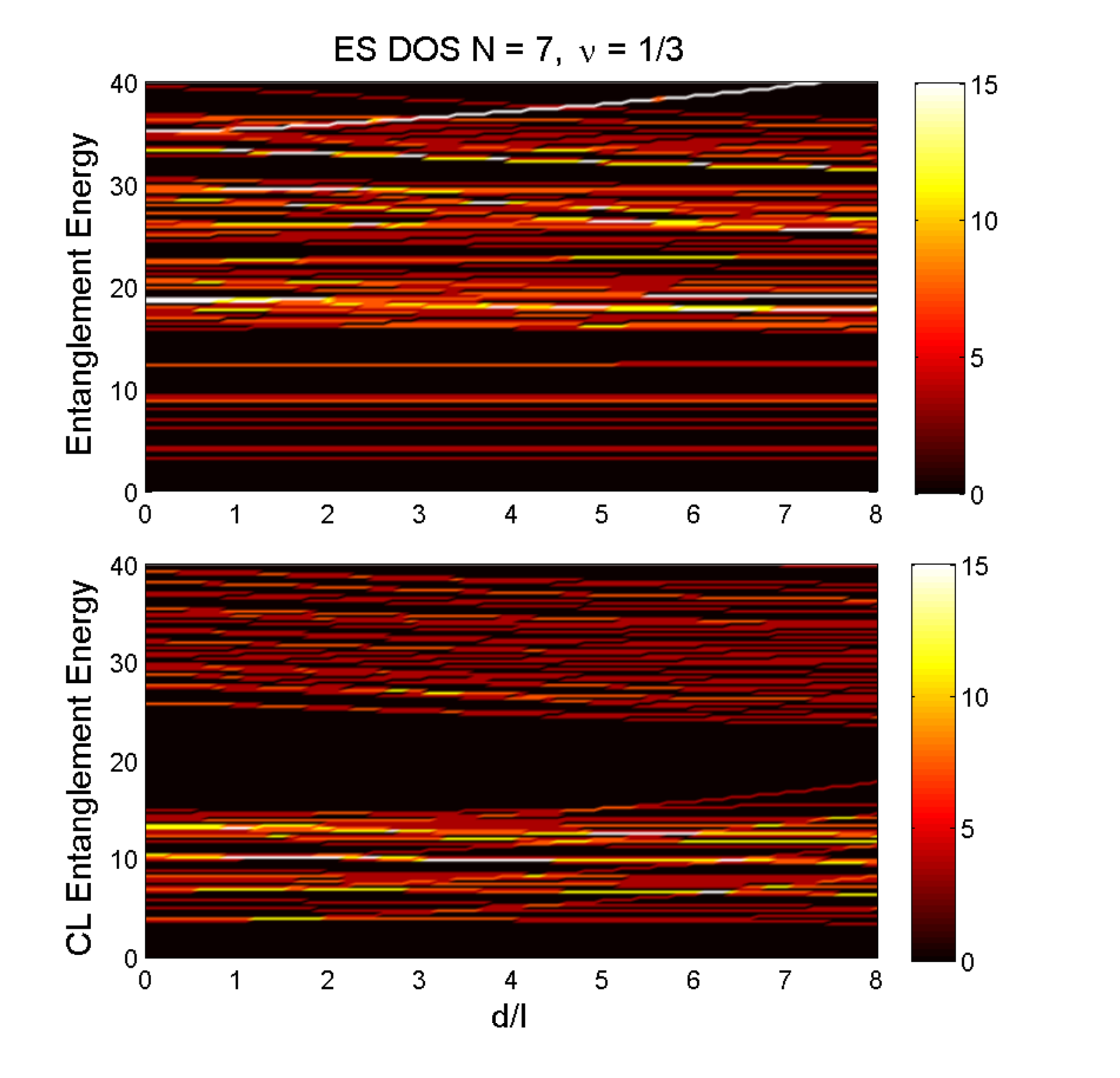}  
\caption{\label{fig:DOS1_3N7} (Color online) Density of entanglement energies before (top panel) and after (bottom panel) the conformal limit for $\nu=1/3$ and particle number $N=7$ as a function of finite layer thickness $d/l$.}
\end{figure}

\begin{figure}
\includegraphics[width=.47\textwidth]{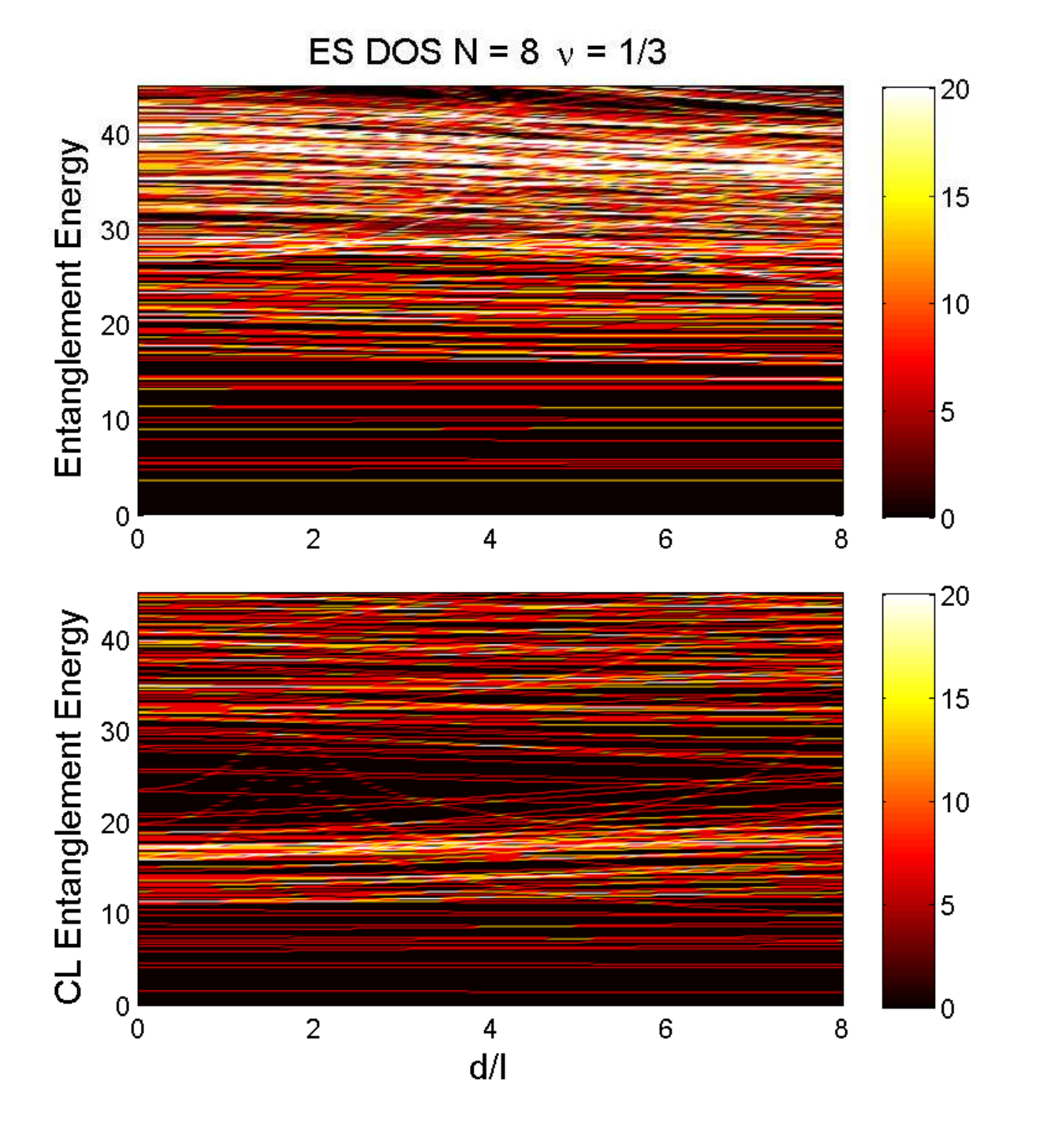}  
\caption{\label{fig:DOS1_3N8} (Color online) Density of entanglement energies before (top panel) and after (bottom panel) the conformal limit for $\nu=1/3$ and particle number $N=8$ as a function of finite layer thickness $d/l$.}
\end{figure}

\begin{figure}
\includegraphics[width=.47\textwidth]{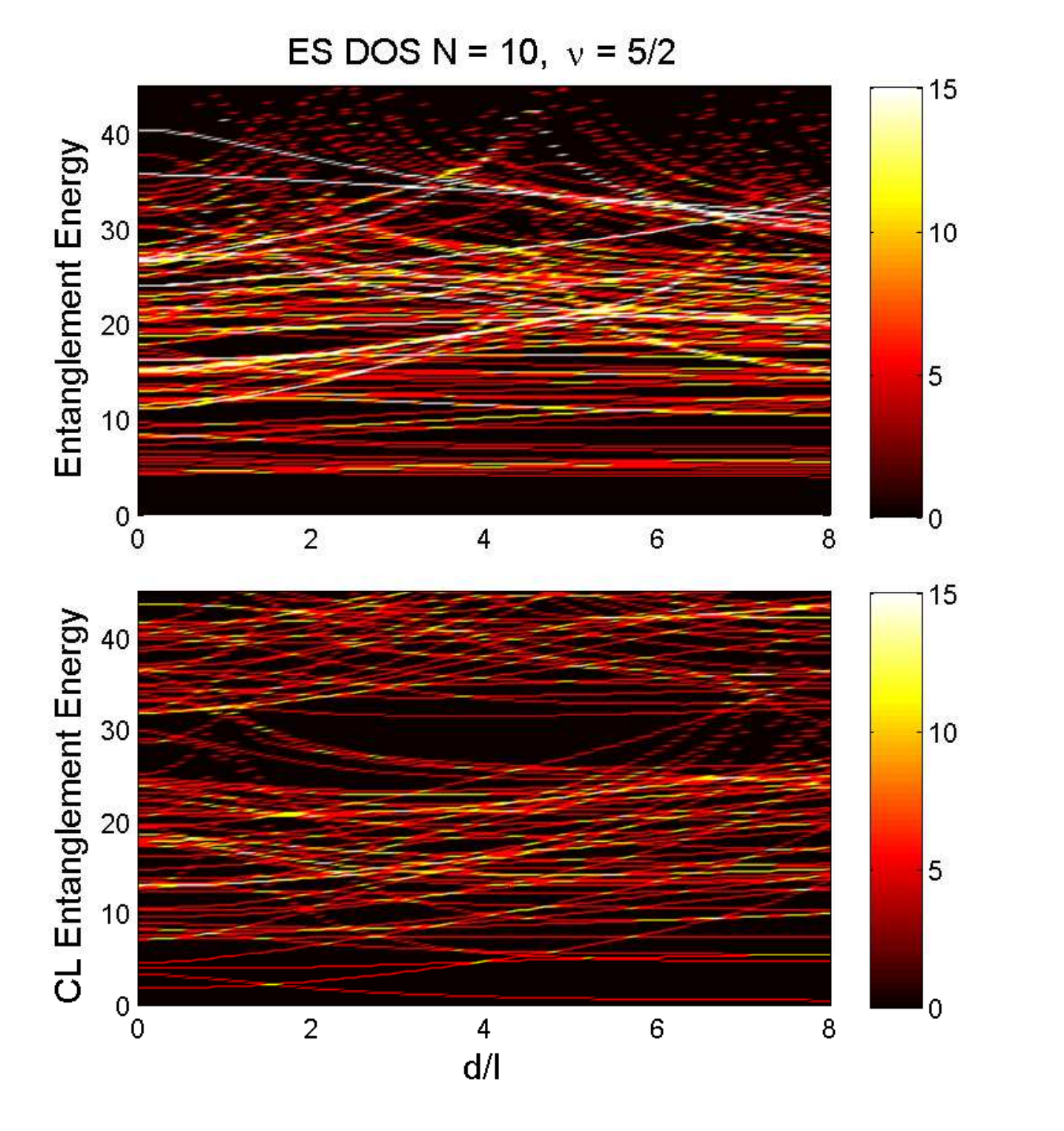}  
\caption{\label{fig:DOS5_2N10} (Color online) Density of entanglement energies before (top panel) and after (bottom panel) the conformal limit for $\nu=1/3$ and particle number $N=7$ as a function of finite layer thickness $d/l$.}
\end{figure}

The plots shown in Fig.~\ref{fig:DOS1_3N7} give the density of states of the ES and CLES for $\nu=1/3$ as a function of finite layer thickness $d/l$ for $N=7$.  In the DOS for the ES before taking the CL, we see sparse low lying states that are separated from a denser cloud of higher states by a series of gaps.  These low lying states are the CFT states from the Li and Haldane conjecture. The states appear, largely to be insensitive to the finite layer thickness.  Turning to the DOS for the CLES, a clear gap is much more evident for the $N=7$ cases.  Here the higher-energy states appear to makeup a wide, low-density band that is well-separated from a low, dense band of states by a gap that decreases with $d/l$.  This case seems to illustrate the effect of using the CL.

For comparison, we provide the DOS results for $N=8$ in Fig.~\ref{fig:DOS1_3N8}.  Here there also appears to be low-lying CFT states in the ES below a high density region of higher energy states.  In the DOS of the CLES, a ``clear'' gap does not appear.  But the low lying band in this case does appear qualitatively similar to the $N=7$ case.  One may possibly associate a ``soft" gap in this case, where a few states appear to be present between two somewhat distinct regions in the DOS.  This ``soft'' gap is qualitatively similar to the ``clear'' gap in the $N=7$ case and it does appear to slightly decrease as a function of $d/l$.  However, it is difficult to distinguish this ``soft'' gap from the other small gaps in the spectrum.

In Fig.~\ref{fig:DOS5_2N10} we provide DOS plots for the $\nu = 5/2$ FQH state for $N=10$ as a function of $d/l$.  In this case,  the ES is especially sensitive to finite layer thickness.  However, we still see a series of small gaps separating thin, dense bands at lower energies.  After taking the CL, a clear gap at finite (non-zero) thickness has a definite peak corresponding to a level crossing.  Below the gap, there appears to be some band crossings as $d/l$ is varied.  

In summary, the DOS of the entanglement spectra (with and without the conformal limit) gives us a general qualitative picture of how the ES evolve with a varying parameter (i.e., the finite layer thickness $d$ in our case).  Thus we expect the DOS of the ES to be a good initial cursory tool in examining topological states with varying parameters.
\section{Conclusions} \label{sec:Conclusion}

In this work we study entanglement in finite sized, quasi-2D FQH states via the entanglement entropy and the entanglement spectrum as a function of the finite layer thickness of the transverse dimension in a realistic FQH system and compare them to the entanglement signatures of the Laughlin and MR Pfaffian model states.  For the Laughlin filling fractions, we find that the EE increases (decreases) with finite layer thickness for $\nu = 1/3$ ($\nu = 7/3$) in the LLL (SLL) with increasing (decreasing) deviation from the EE of the Laughlin model state.  However the EE in the SLL reaches an asymptotic value larger than the EE of the Laughlin state, possibly suggesting the $\nu = 7/3$ state is modeled by different physics than the Laughlin state. Similar behavior is also seen in the entanglement gaps of the ES for the Laughlin filling fractions.  Here we find that the entanglement gaps decrease with finite layer thickness for the Laughlin filling fractions in the LLL.  But in the SLL, the behavior of the entanglement gaps depend on the ``depth'' of the gap.  These results suggest that the Laughlin FQH states ``weaken" with increasing thickness in the LLL, which is consistent with previous work on quasi-2D FQH states \cite{PetersonPRL08,Peterson08}, but in the SLL, other physics beyond just the Laughlin state alone is needed to describe the FQH state.    The LL dependence of the finite thickness effect at half-filling differs slightly.  The EE of the $\nu = 1/2$ state in the LLL is largely insensitive to the finite layer thickness in contrast to that of the SLL $\nu = 5/2$ state where the EE has a local minima that approaches the EE of the MR Pfaffian at finite $d/l$.  This qualitative behavior is also seen in the entanglement gaps of the ES for half-filled LLs.  For $\nu = 5/2$ in the SLL, we see peaks (local maxima) in the entanglement gaps at finite thickness, suggesting the $\nu = 5/2$ is more ``MR Pfaffian-like" at an optimal thickness, which, again is consistent with previous work\cite{PetersonPRL08,Peterson08} and strongly suggests the $\nu = 5/2$ state is, indeed, MR Pfaffian.  In contrast the entanglement gaps of the ES for the $\nu = 1/2$ state suggest that it is not modeled by the MR Pfaffian.  Thus, the entanglement gaps in the ES allows us to differentiate the $\nu = 1/2$ and $\nu = 5/2$ states, which we could not definitively establish with the EE or the overlap calculations.  Of course, we must be cautious with these results since the calculated entanglement gaps made use of only a few ``Virasoro levels" in the low-lying CFT due to finite-size effects.  Assuming the Li and Haldane conjecture to be correct, we can only say that we have observed the Laughlin and MR Pfaffian signatures up to a few ``Virasoro levels".

We also investigate the conformal limit of the entanglement spectrum which is conjectured to remove curvature in the spectrum due to finite size effects and allow the use of the entire spectrum to determine the topological signature of the state.  Our results on the conformal limit, however are inconsistent between varying system sizes and are difficult to interpret.   This appears to be due to our choice of using planar pseudopotentials rather than spherical pseudopotentials in obtaining the FQHE ground states.  In the appendix we examine this choice by comparing the entanglement spectra of ground states obtained by using either spherical or planar pseudopotentials at $d=0$ and observe that the conformal limit can be affected by components of the ground state that have exponentially small contributions and, therefore, are sensitive to minor details in the interaction (such as the difference between planar and spherical pseudopotentials).  Thus the presence of the entanglement gap in the conformal limit is sensitive to certain details in the effective interaction that may not be relevant in determining the topological features of the state. Further work using much larger system sizes would be necessary to resolve this issue which is well beyond the scope of the current work.

We have also introduced the notion of entanglement density of states as a method for examining the idea of an entanglement gap without an explicit reliance on a model wavefunction.  Although, far from definitive, the entanglement DOS  suggests itself as a powerful tool to determine the topological nature of a particular ground state.  Our detailed numerical study establishes the entanglement DOS to be a useful quantity underlying topological FQHE particularly in the context of finite size numerical calculations.

It is interesting to observe that the entanglement measures give results similar to those obtained with overlaps in Refs.~\onlinecite{PetersonPRL08, Peterson08}.  Whereas the overlap is a simple measure of how well a numerically obtained ground state matches a particular model state (e.g. the Laughlin state or the MR Pfaffian state), the entanglement measures (in particular, the ES) is a more general measure of how well a state fits a suspected CFT (i.e. universality class) that describes the model state.  Therefore, it can be said that these results confirm the conclusions in Refs.~\onlinecite{PetersonPRL08, Peterson08} in a more general sense in respect to the Laughlin and MR Pfaffian CFTs.  However, we must be cautious in this generalization given that we have only observed the Laughlin and MR Pfaffian signature up to a few ``Virasoro levels" and different theories can result in the same low-level structure in the ES~\cite{Turner}.  More work is necessary to understand how well entanglement measures can definitely identify universality classes in finite systems.

In interpreting our results and conclusions, one may wonder about the importance of finite size effects on our numerical diagonalization.  The possible limitations associated with finite-size effects are of course always present in any exact diagonalization  study of any FQHE system, and the possibility that some of the conclusions are affected by finite size effects can never be ruled out even if the calculations are carried out on systems much larger than what we use in this work, since in the end any statement about an experimental system based on calculations performed on few-particle systems is always subject to an extrapolation to the thermodynamic limit.  We believe that all our conclusions regarding the importance of finite quasi-2D thickness effect on the FQHE entanglement spectra are valid independent of the rather modest size of our finite system diagonalization study because earlier work~\cite{PetersonPRL08, Peterson08} clearly established, when compared with calculations ~\cite{StorniPRL,Feiguin08}  carried out on much larger systems, that the system size we use in this work, namely N=8, is certainly adequate in making qualitatively correct conclusions about the SLL FQHE.  Our goal in this paper has been to study as many FQHE states as feasible as a function of the quasi-2D layer thickness in depth, thus necessarily (due to the computational time restrictions) limiting our system size to N=8 which should be adequate.  Nevertheless, we feel that future work should explore larger system size diagonalization in order to study the finite-thickness effect on the entanglement spectra of various FQHE states.
  
We also note that there are various alternatives to the infinite square well effective potential in examining the finite thickness effect.  The Zhang-Das Sarma potential is likely the most well-known and oft-used alternative in this regard.  We choose to focus on the infinite square well instead of the Zhang-Das Sarma potential because the infinite square well is more closely aligned to realistic quasi-2D systems.  Moreover many observables calculated with the Zhang-Das Sarma potential besides the overlap have been shown to give similar results compared to the infinite square well.  Therefore we expect the two potentials to give similar results with entanglement measures as well, but since investigating the nature of these potentials is not a goal of the present study, we leave explicit verification of this assertion for a future work. 
  
It should be noted that the concept of entanglement spectra (or for that matter, the entanglement entropy itself) has no direct experimental or observational implications since it cannot be directly measured.  The concept is useful conceptually and theoretically in ascertaining the quantum topological nature of a particular interacting Hamiltonian, and in that sense its experimental consequences are indirect since the topological nature of a system has obvious experimental consequences.  We have investigated in this work the utility of the concept of the entanglement spectra in ascertaining the underlying topological nature of realistic FQHE states as a function of the quasi-2D layer thickness, finding that the entanglement spectrum provides results consistent with what has earlier been established in the literature based on the wavefunction overlap studies.  Our very detailed study also indicates that at this stage of theoretical development, the entanglement approach is perhaps no more predictive in providing experimental implications of various FQHE states than what is already available in the literature based on the direct wavefunction overlap studies.  Further work would be necessary to see if the entanglement approach has some specific advantages in predicting experimental properties of FQHE states not already apparent in wavefunction-based analyses.

In conclusion, we have extended the concept of topological entanglement spectra and entanglement gaps to finite-thickness FQH systems by calculating the FQHE topological properties systemically as a function of finite thickness of the quasi-2D systems, establishing in the process that the FQHE entanglement measures calculated as a function of system thickness are completely consistent with the results obtained earlier in the literature using wavefunction overlap calculations.  While our work establishes various entanglement measures as important theoretical quantities classifying FQHE, more work will be necessary to understand the finite size aspects of entanglement spectra and entanglement gaps in the context of realistic fractional quantum Hall systems.  Although it is gratifying that the qualitative conclusions of our entanglement-measure-based results in this work are completely consistent with earlier FQHE results obtained on the basis of wavefunction overlap calculations, it remains to be seen whether the entanglement-measure based probes have more predictive power regarding the nature of FQHE than the wavefunction-overlap based probes or it is simply a deeper way of looking at the same physics with no obvious additional implications for the experimental occurrence of FQHE.

\appendix
\section{Planar vs spherical pseudopotentials at $d=0$} \label{sec:appendix}

The analysis presented above is based on ground state wavefunctions obtained by diagonalizing the quasi-2D Coulomb potential in a spherical geometry.  However the Haldane pseudopotentials used to construct the Hamiltonian are derived from a infinite planar geometry rather than a spherical geometry.  We choose to use planar rather than spherical psuedopotentials because (i) the effective Coulomb potential in a quasi-2D system is more naturally obtained in the infinite planar geometry and (ii) we expect the spherical and planar pseudopotentials to be indistinguishable in the thermodynamic limit.  Moreover given the mostly qualitative nature involved in studying entanglement spectra, we expected this choice to make little difference in the results.  Nevertheless, there are cases under study where this choice matters.  The goal of this appendix is to highlight some of these cases.  We show that for $d=0$, the low energy spectrum in the entanglement spectra are qualitatively similar between ground states obtained from either spherical or planar pseudopotentials, but higher energy spectra can can differ in some cases.  This difference does not change the qualitative conclusions drawn from the low energy spectra, but when we consider the conformal limit which looks for a full entanglement gap, the difference can lead to different conclusions (in particular, the case of $\nu = 1/3$ with $N=8$). We leave the comparison of cases with  $d>0$ and larger $N$ for future work.

\begin{table}
\begin{tabular*}{.5\textwidth} { c | c | c | c | c}
\hline 
$N$ & $\nu$ & $\left\langle\Psi_{\mathrm{sphere}}|\Psi_{\mathrm{plane}}\right\rangle$  & $\left\langle\Psi_{\mathrm{sphere}}|\Psi_{\mathrm{model}}\right\rangle$ & $\left\langle\Psi_{\mathrm{plane}}|\Psi_{\mathrm{model}}\right\rangle$ \\
\hline \hline
6		& 1/3 & 0.9988 & 0.9964  & 0.9921 \\ \hline
6		&	7/3	& 0.9480 & 0.5285  & 0.7369 \\ \hline
7		& 1/3 & 0.9999 & 0.9964  & 0.9952 \\ \hline
7   & 7/3 & 0.8648 & 0.6071  & 0.8737 \\ \hline
8   & 1/3 & 0.9996 & 0.9954  & 0.9954 \\ \hline
8		& 7/3 & 0.9675 & 0.5719  & 0.7441 \\ \hline
8		& 1/2 & 0.9978 & 0.9213  & 0.8953 \\ \hline
8   & 5/2 & 0.9688 & 0.8674  & 0.9639 \\ \hline
10  & 1/2 & 0      & 0.8891  & 0      \\ \hline
10  & 5/2 & 0.9720 & 0.8376  & 0.9342 \\
\hline  
\end{tabular*}
\caption{\label{tab:overlaps} Overlap integrals between 1) the exact ground state wavefunction using spherical ($\left|\Psi_{\mathrm{sphere}}\right\rangle$) and planar Haldane pseudopotentials ($\left|\Psi_{\mathrm{plane}}\right\rangle$), and 2) the overlap between the Laughlin or Pfaffian wavefunction ($\left|\Psi_{\mathrm{model}}\right\rangle$ and the exact ground state wavefunction using spherical or planar pseudopotentials. }
\end{table}

In Table \ref{tab:overlaps} we provide several overlap calculations between exact ground states at $d=0$ obtained using either spherical or planar pseudopotentials.  In column 3 of the table, we see that the overlap between the ground states from the spherical and planar cases is generally high. The notable exception is the case when $N=10$ and $\nu=1/2$ where the overlap is $0$.  In this case the ground state obtained with the planar pseudopotentials possesses a different symmetry compared to the ground state of the spherical case, which leads to a vanishing overlap.  Excluding these, columns 4 and 5 of the table show that the overlap between the spherical and planar ground states with the model Laughlin or MR Pfaffian states are qualitatively similar.  

\begin{figure*}
\includegraphics[width=.99\textwidth]{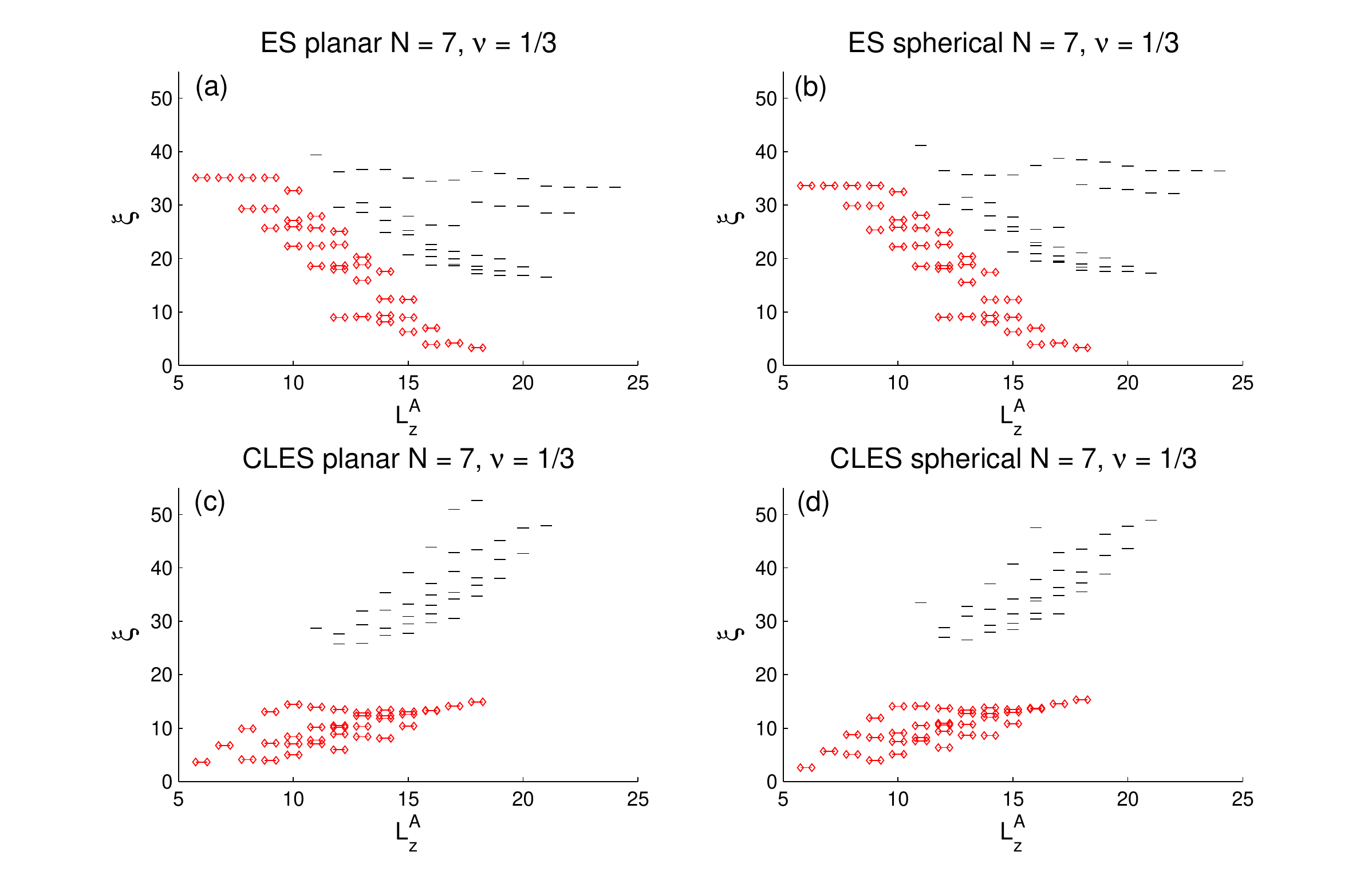}  
\caption{\label{fig:SphereVsPlanarN7nu1_3} (Color online) Entanglement spectra and conformal limit entanglement spectra of the exact FQHE ground state for $N=7$ at filling fraction $\nu=1/3$ obtained with either planar or spherical Haldane pseudopotentials at $d=0$.  CFT states associated with the Laughlin model wavefunction are marked.}
\end{figure*}

\begin{figure*}
\includegraphics[width=.99\textwidth]{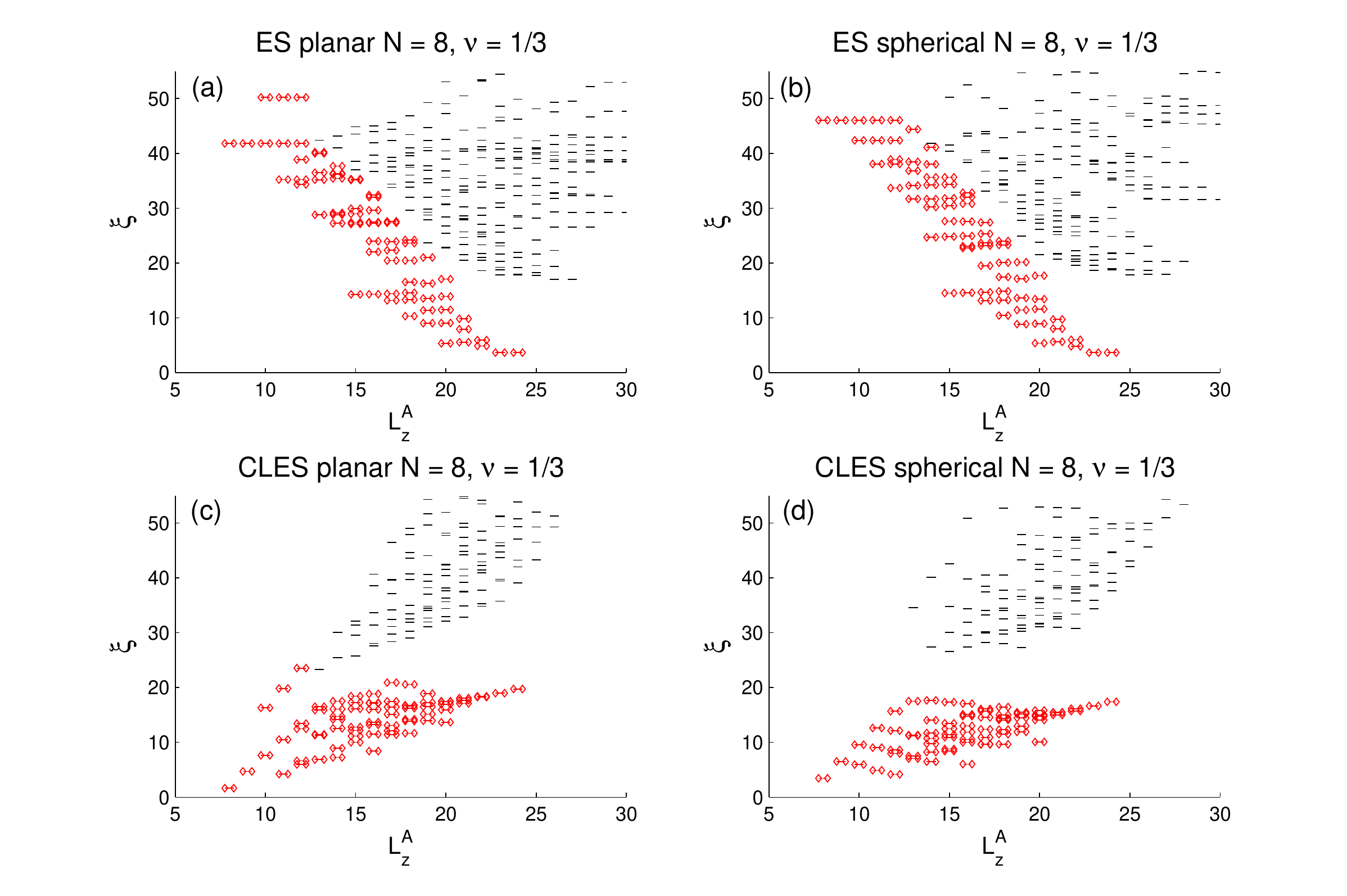}  
\caption{\label{fig:SphereVsPlanarN8nu1_3} (Color online) Entanglement spectra and conformal limit entanglement spectra of the exact FQHE ground state for $N=8$ at filling fraction $\nu=1/3$ obtained with either planar or spherical Haldane pseudopotentials at $d=0$.  CFT states associated with the Laughlin model wavefunction are marked.}
\end{figure*}

\begin{figure*}
\includegraphics[width=.99\textwidth]{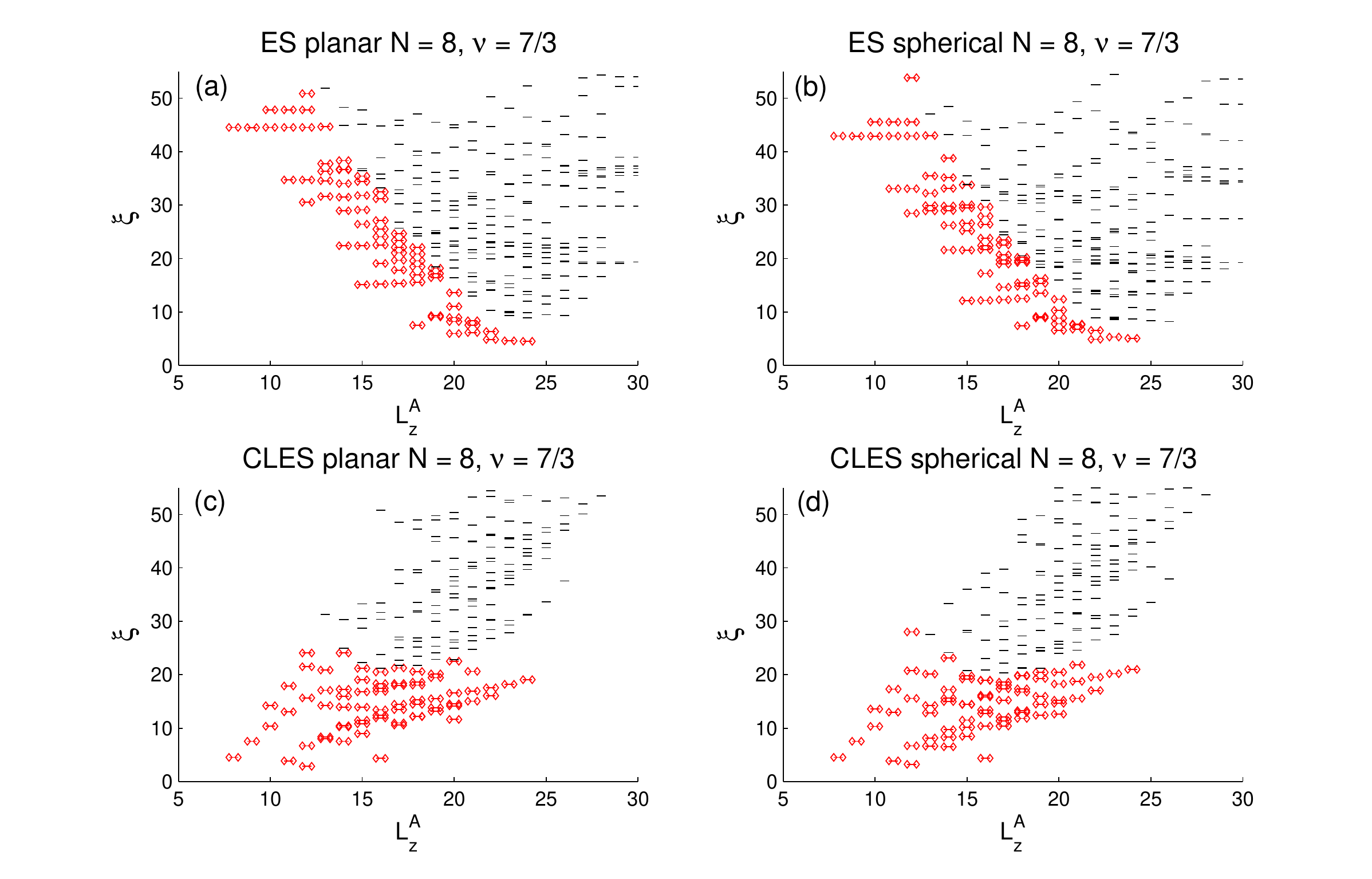}  
\caption{\label{fig:SphereVsPlanarN8nu7_3} (Color online) Entanglement spectra and conformal limit entanglement spectra of the exact FQHE ground state for $N=7$ at filling fraction $\nu=7/3$ obtained with either planar or spherical Haldane pseudopotentials at $d=0$.  CFT states associated with the Laughlin model wavefunction are marked.}
\end{figure*}

\begin{figure*}
\includegraphics[width=.99\textwidth]{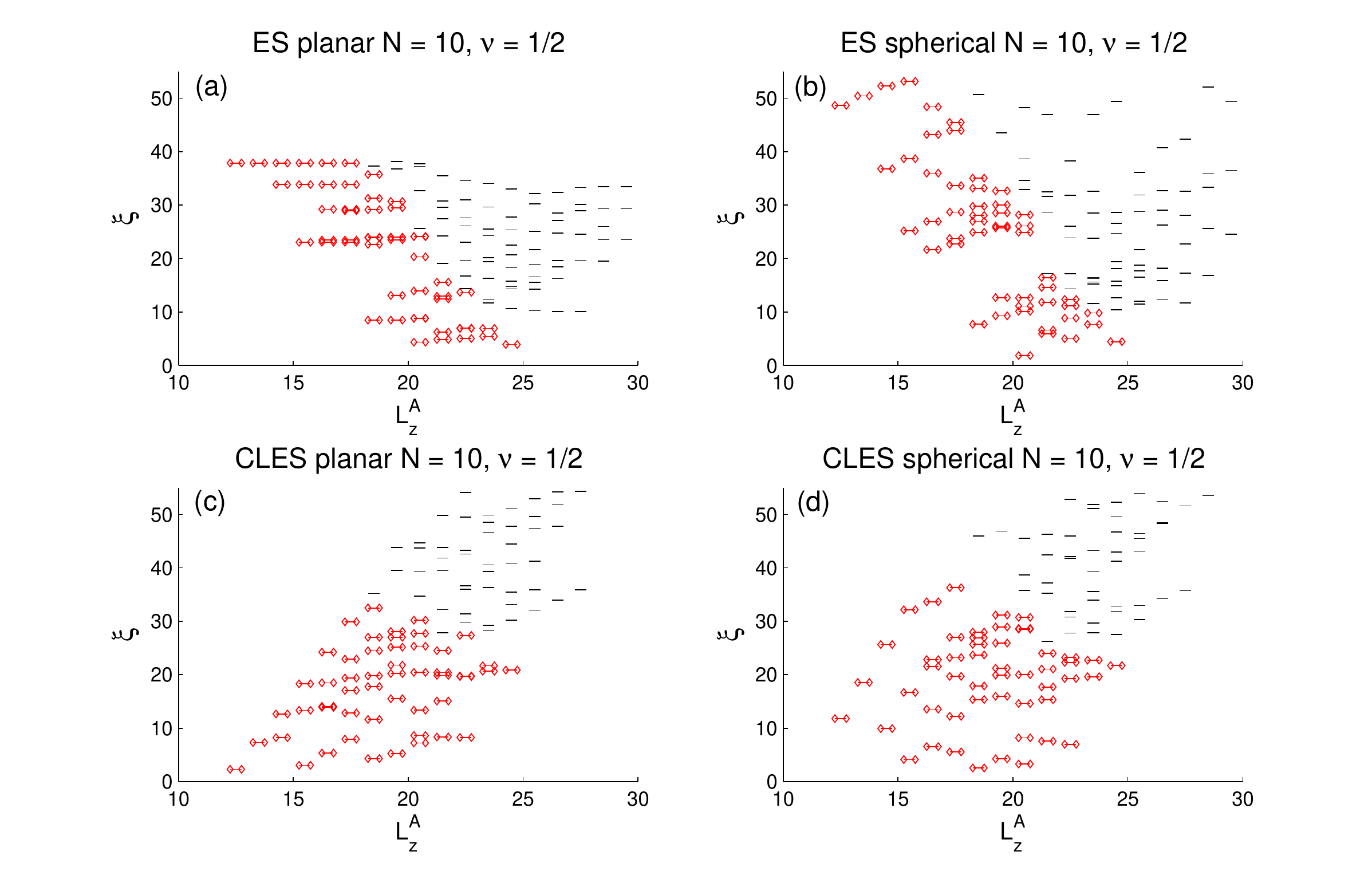}  
\caption{\label{fig:SphereVsPlanarN10nu1_2} (Color online) Entanglement spectra and conformal limit entanglement spectra of the exact FQHE ground state for $N=10$ at filling fraction $\nu=1/2$ obtained with either planar or spherical Haldane pseudopotentials at $d=0$.  CFT states associated with the MR Pfaffian model wavefunction are marked.}
\end{figure*}

\begin{figure*}
\includegraphics[width=.99\textwidth]{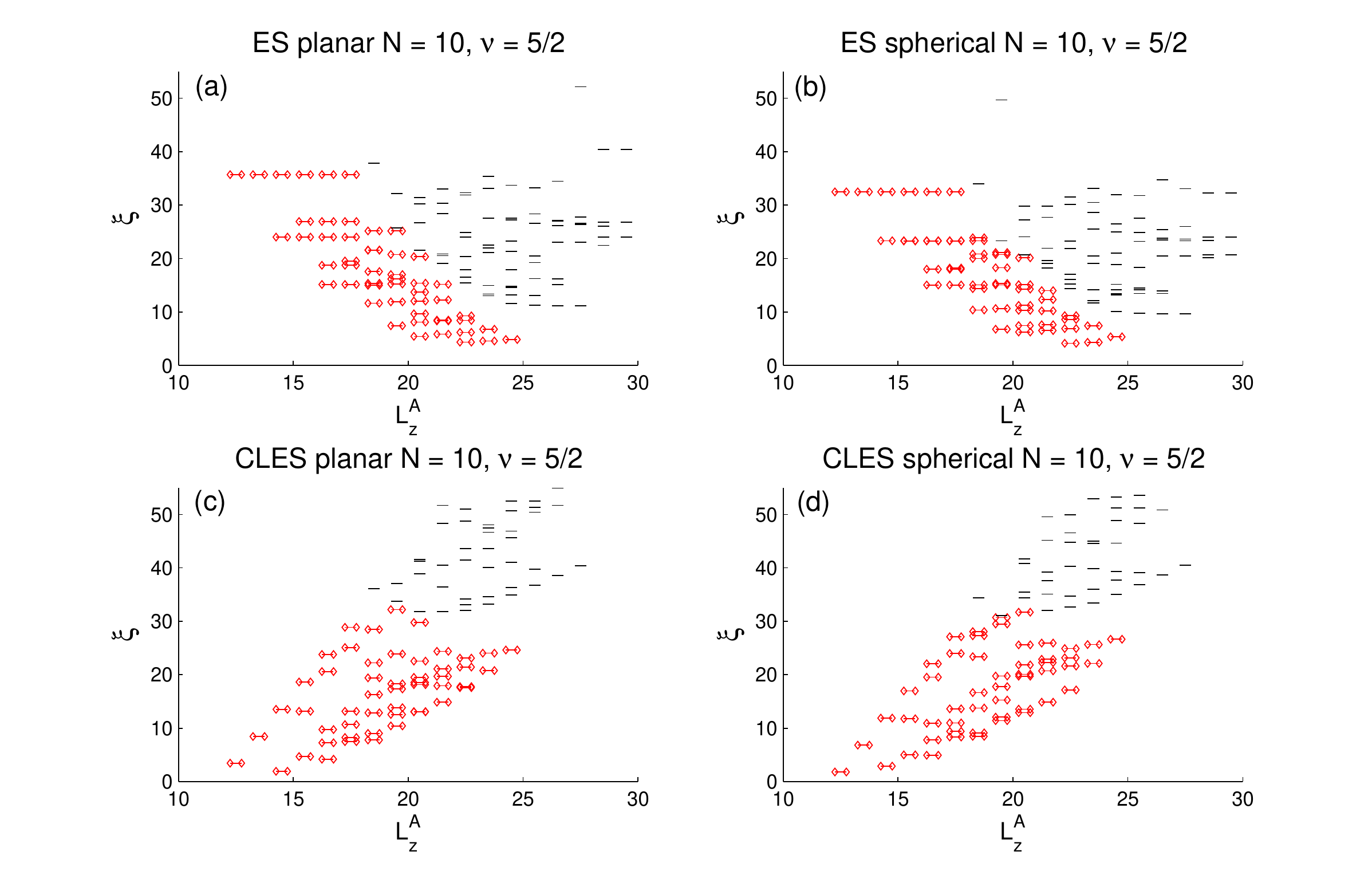}  
\caption{\label{fig:SphereVsPlanarN10nu5_2} (Color online) Entanglement spectra and conformal limit entanglement spectra of the exact FQHE ground state for $N=10$ at filling fraction $\nu=5/2$ obtained with either planar or spherical Haldane pseudopotentials at $d=0$.  CFT states associated with the MR Pfaffian model wavefunction are marked.}
\end{figure*}

We now turn our attention to the entanglement spectra and how they may differ with choice of pseudopotentials.  ES (with and without the conformal limit) for the exact ground state of the FQHE state at $N=7$ and $\nu=1/3$ using spherical and planar pseudopotentials are given in fig. \ref{fig:SphereVsPlanarN7nu1_3}.  In the figure, we see that the ES with planar pseudopotentials (Fig. \ref{fig:SphereVsPlanarN7nu1_3} a) is qualitatively similar to the spectra obtained with spherical pseudopotentials (Fig. \ref{fig:SphereVsPlanarN7nu1_3} b).  The same can also be said with the ES in the conformal limit between the planar case (Fig. \ref{fig:SphereVsPlanarN7nu1_3} c) and the spherical case (Fig. \ref{fig:SphereVsPlanarN7nu1_3} d).  Thus, given the results in Fig. \ref{fig:SphereVsPlanarN7nu1_3}, we would expect that the choice of pseudopotenials makes little difference in obtaining a qualitative understanding of the ES in this case.

Fig. \ref{fig:SphereVsPlanarN8nu1_3} compares the ES of the FQHE state at filling fraction $\nu=1/3$ with $N=8$.  In this case we see that in the ES before the conformal limit (Fig. \ref{fig:SphereVsPlanarN8nu1_3} a and b), the low energy spectra are qualitatively similar between the planar and spherical cases.  The higher energy spectra in the ES, however, show notable differences with the planar case having a few CFT levels at much higher energy compared to the spherical case.  In the conformal limit, these higher energy CFT levels lead to a vanishing entanglement gap in the conformal limit for the planar case (Fig. \ref{fig:SphereVsPlanarN8nu1_3} c) compared to the spherical case (Fig. \ref{fig:SphereVsPlanarN8nu1_3} d) where there is a full entanglement gap.  These are the same ``spurious'' levels identified earlier in section \ref{sec:CL}. These results suggest that the vanishing minimal gap seen in Fig. \ref{fig:CLES1_3} is due to our choice of planar rather than spherical Haldane pseudopotentials.  This may seem surprising given the large overlaps seen in Table \ref{tab:overlaps}. However the states associated with the the higher energy CFT levels have exponentially small contributions to the ground state wavefunction, and thus contribute little to the overlap.  We might also expect these states to be more sensitive to certain quantitative details of the potential that do not affect the qualitative picture of the FQHE ground state (e.g. values of $V_m$ for ``large'' $m$).  Thus when taking the conformal limit, the choice of pseudopotential may matter in some cases in order to observe a full entanglement gap.  But a qualitative understanding can still be gleaned from the planar case since there does appear to be two distinct regions in the CLES that we can identify, at least qualitatively, as CFT and non-CFT levels.

We now compare the spherical and planar pseudopotentials in the SLL with $\nu=7/3$.  Fig. \ref{fig:SphereVsPlanarN8nu7_3} gives the ES (\ref{fig:SphereVsPlanarN8nu7_3} a and b) and CLES (\ref{fig:SphereVsPlanarN8nu7_3} c and d) for the $\nu=7/3$ FQHE state obtained with either planar or spherical pseudopotentials with $N=8$.  The planar and spherical cases are qualitatively similar in both the ES and CLES and both suggest that the Laughlin wavefunction may not describe this state, as discussed in sections \ref{sec:EE}, \ref{sec:ES} and \ref{sec:CL}.

Results for the even denominator filling fraction $\nu=1/2$ with $N=10$ are given in Fig. \ref{fig:SphereVsPlanarN10nu1_2}.  In this case, the planar results (\ref{fig:SphereVsPlanarN8nu1_3} a and c) differ considerably from that of the spherical case (\ref{fig:SphereVsPlanarN8nu1_3} b and d).  This is not surprising since the overlap between these two states given in table \ref{tab:overlaps} vanishes.  However, it appears that neither state is consistent with the MR Pfaffian.

Comparison of FQHE ground states obtained with planar and spherical pseudopotentials for the $\nu=5/2$ state with $N=10$ is given in Fig. \ref{fig:SphereVsPlanarN10nu5_2}.  Similar to the $\nu=1/3$ case, the low energy spectra in the ES (\ref{fig:SphereVsPlanarN10nu5_2} a and b) are qualitatively similar between the two cases.  The higher energy levels in the spectra do differ, but the CLES (\ref{fig:SphereVsPlanarN10nu5_2} c and d) does appear to give the same qualitative picture.  Recall that in section \ref{sec:CL}, the minimal gap for this case becomes non-zero only at finite $d$ for the planar case.  We would expect a similar result to occur using the spherical psuedopotentials.  Verification of this is left for a future work.

\begin{acknowledgments}
This work is supported by Microsoft Q and DARPA QUEST.
\end{acknowledgments}

\begin{thebibliography}{89}
\bibitem[{foo()}]{footnote}
\bibinfo{note}{Refs.~\onlinecite{PetersonPRL08,Peterson08} also showed the
  emergence of the so-called three-fold topological degeneracy which is a
  signature of the MR Pfaffian state on the torus geometry with the inclusion
  of finite thickness.}
\end{thebibliography}

\end{document}